\documentclass{aastex631}
\usepackage{upgreek}
\usepackage{newtxtext,newtxmath}
\usepackage[normalem]{ulem}

\newcommand{\SNR}{\mathrm{S}/\mathrm{N}}
\newcommand{\SNRp}{\mathrm{S}_p/\mathrm{N}_p}
\newcommand{\vlos}{v_\mathrm{los}}
\newcommand{\Teff}{T_\mathrm{eff}}
\newcommand{\Neff}{N_\mathrm{eff}}
\newcommand{\neff}{n_\mathrm{eff}}
\newcommand{\logg}{\log g}
\newcommand{\MH}{\left[ \mathrm{M} / \mathrm{H} \right]{}}
\newcommand{\FeH}{\left[ \mathrm{Fe} / \mathrm{H} \right]{}}
\newcommand{\aM}{\left[ \upalpha / \mathrm{M} \right]{}}

\newcommand{\given}{\,\vert\,}
\newcommand{\dd}{\,\mathrm{d}}

\def\vertical_distance{7pt}

\newcommand{\mdet}[1]{\big\vert \,#1\, \big\vert}
\newcommand{\bigmdet}[1]{\Bigg\vert \,#1\, \Bigg\vert}
\newcommand{\T}{\mathrm{T}}
\newcommand{\bA}{\mathbf{A}}
\newcommand{\bB}{\mathbf{B}}

\newcommand{\bD}{\mathbf{D}}
\newcommand{\bF}{\mathbf{F}}
\newcommand{\bG}{\mathbf{G}}
\newcommand{\bS}{\mathbf{S}}
\newcommand{\bK}{\mathbf{K}}
\newcommand{\bM}{\mathbf{M}}
\newcommand{\bMi}{\mathbf{M}^{-1}}
\newcommand{\bU}{\mathbf{U}}
\newcommand{\ba}{\mathbf{a}}
\newcommand{\bb}{\mathbf{b}}

\newcommand{\bd}{\mathbf{d}}
\newcommand{\bg}{\mathbf{g}}
\newcommand{\bs}{\mathbf{s}}
\newcommand{\bphi}{\boldsymbol\varphi}

\newcommand{\bchi}{\boldsymbol{\chi}}
\newcommand{\bchiI}{\boldsymbol\chi_0^{-1}}

\newcommand{\bSigma}{\boldsymbol\Sigma}

\begin{document}

\title{Uncertainty of line-of-sight velocity measurement of faint stars from low and medium resolution optical spectra}

\correspondingauthor{László Dobos}
\email{dobos@jhu.edu}

\author[0000-0001-7679-9478]{László Dobos}
\affiliation{Department of Physics \& Astronomy, The Johns Hopkins University, Baltimore, MD 21218, USA}

\author{Alexander S. Szalay}
\affiliation{Department of Physics \& Astronomy, The Johns Hopkins University, Baltimore, MD 21218, USA}

\author{Tamás Budavári}
\affiliation{Department of Applied Mathematics \& Statistics, Johns Hopkins University, Baltimore, MD 21218, USA}
\affiliation{Department of Computer Science, Johns Hopkins University, Baltimore, MD 21218, USA}
\affiliation{Department of Physics \& Astronomy, The Johns Hopkins University, Baltimore, MD 21218, USA}

\author{Evan N. Kirby}
\affiliation{Department of Physics, University of Notre Dame, Notre Dame, IN 46556, USA}

\author{Robert H. Lupton}
\affiliation{Department of Astrophysical Sciences, Princeton University, Peyton Hall, Princeton, NJ 08544, USA}

\author[0000-0002-4013-1799]{Rosemary F.G. Wyse}
\affiliation{Department of Physics \& Astronomy, The Johns Hopkins University, Baltimore, MD 21218, USA}

\begin{abstract}
Massively multiplexed spectrographs will soon gather large statistical samples of stellar spectra. The accurate estimation of uncertainties on derived parameters, such as line-of-sight velocity $\vlos$, especially for spectra with low signal-to-noise ratios, is paramount. We generated an ensemble of simulated optical spectra of stars as if they were observed with low- and medium-resolution fiber-fed instruments on an 8-meter class telescope, similar to the Subaru Prime Focus Spectrograph, and determined $\vlos$ by fitting stellar templates to the simulations. We compared the empirical errors of the derived paramaters -- calculated from an ensemble of simulations -- to the asymptotic error determined from the Fisher matrix, as well as from Monte Carlo sampling of the posterior probability. We confirm that the uncertainty of $\vlos$ scales with the inverse square root of $\SNR$, but also show how this scaling breaks down at low $\SNR$ and analyze the error and bias caused by template mismatch. We outline a computationally optimized algorithm to fit multi-exposure data and provide the mathematical model of stellar spectrum fitting that maximizes the so called \textit{significance}, which allows for calculating the error from the Fisher matrix analytically. We also introduce the \textit{effective line count}, and provide a scaling relation to estimate the error of $\vlos$ measurement based on the stellar type. Our analysis covers a range of stellar types with parameters that are typical of the Galactic outer disk and halo, together with analogs of stars in M31 and in satellite dwarf spheroidal galaxies around the Milky Way.

\end{abstract}


\section{Introduction}

To decipher the assembly history of the Milky Way and neighboring galaxies, Galactic Archaeology relies on measuring the dynamical and chemical properties of individual stars throughout the outer disk and the halo of the Galaxy, as well as stars in satellite galaxies and M31. The low number density of target stars, the requirement of sufficiently large sample sizes for statistical analysis, and the large celestial area to be covered make wide field-of-view 4-~and 8-meter class telescopes with low and medium resolution optical fiber-fed spectrographs -- such as the Subaru Prime Focus Spectrograph (PFS) \citep{Sugai2014SPIE.9147E..0TS, Takada2014PASJ...66R...1T} or the Dark Energy Spectroscopic Instrument (DESI) \citep{DESI2022AJ....164..207D, Dey2023ApJ...944....1D} -- the best available tools for the observations. Even with these advanced instruments, constraints on observational time limit the achievable signal-to-noise ratio to a range of $\SNR = 5$-$100$ per resolution element, with the large majority of the targets observed at $\SNR < 15$. Typically, a constraint on the maximally acceptable uncertainties for measured velocities $\sigma(\vlos)$ and metallicities $\sigma(\FeH)$ drive the requirements on signal-to-noise for faint stars. Measuring the abundances of most individual elements is typically reserved for brighter stars with higher signal-to-noise ratios and higher resolution instruments.

In this study, our primary focus is on characterizing $\sigma(\vlos)$, the uncertainty in line-of-sight velocity $\vlos$, based on many repeated realizations of simulated stellar spectra. We characterize the error as a function of spectral type and a wide range of flux signal-to-noise ratios. Understanding the sources of uncertainty in line-of-sight velocity determination and precise estimation of the error is essential to investigate the velocity distributions of dynamically cold systems such as stellar streams and ultra-faint dwarf spheroidal satellite galaxies and tackle problems such as the radial mass profile of dark matter halos in dwarf spheroidal galaxies, as overestimating the uncertainties can directly lead to underestimating the velocity dispersion and \textit{vice versa} \citep{Koposov2011ApJ...736..146K}. Pessimistic error estimates lead to underestimation of the velocity dispersion, especially when the velocity dispersion is comparable to the measurement error. To compile a statistically significant sample, the low potential target density of nearby dwarf spheroidal galaxies might demand a fainter magnitude limit to the survey because targeting a large number of faint stars with less precise line-of-sight velocities -- but with well understood errors -- might have better constraining power on the radial dark matter profile than brighter but smaller samples.

It has been established that the close binary fraction in the stellar populations of the Milky Way increases with decreasing metallicity \citep{Badenes2018ApJ...854..147B, Moe2019ApJ...875...61M}, and a similar dependence is inferred for stellar populations in dwarf spheroidal galaxies, from the frequency of blue stragglers \citep{Wyse2020}. The majority of targets in stellar spectroscopic surveys of Local Group galaxies are expected to be in multiple systems. When the brightnesses of the two stars are comparable, spectroscopic binaries have a significant effect on the uncertainty of $\vlos$ measurements when the multiplicity is not accounted for. In the case of evolved (giant) primaries, the contribution of the companion (dwarf) star to the spectrum is usually negligible. However, at the distance of extragalactic targets, such as red giants in M31 and M33, the stellar surface density can be high enough that the probability that two stars are aligned along the line-of-sight becomes significant \citep{Kounkel2021AJ....162..184K}. Even though these are not gravitationally bound systems, the effect of this multiplicity on the $\vlos$ measurement is similar to that of binaries. The effect of unresolved binaries on the $\vlos$ measurement is an important topic, but it is beyond the scope of the present paper and we defer further discussion to a future study. For a review of currently available template-based and model-independent methods for disentangling spectroscopic binaries from multi-epoch, low resolution spectra, we refer the reader to a recent study by \cite{Seeburger2024MNRAS.530.1935S}.

We simulate spectroscopic observations of single stars with a wide range of flux signal-to-noise ratios, for a broad range of stellar types. We model both the random and systematic observational effects with software based on the spectroscopic Exposure Time Calculator (ETC) originally developed by \citet{Hirata2012arXiv1204.5151H} for the Roman Space Telescope\footnote{Formerly WFIRST.} and adapted to the Subaru PFS by \cite{ETCSubaru}.

We determine $\vlos$ and its measurement error with several methods. We compare the empirical uncertainty of $\vlos$, determined from an ensemble of simulations, to those that one can derive from individual spectra via mathematical methods. These methods include the evaluation of the Fisher matrix of a maximum likelihood or maximum significance fit using a single model spectrum as template. In addition to fitting a single template to determine $\vlos$, we also optimize for the atmospheric parameters by interpolating the templates on a synthetic stellar spectrum grid. Although this allows for infering the stellar parameters, including effective temperature, metallicity and surface gravity, in the present study, we restrict our focus to the uncertainties of the line-of-sight velocity. Nevertheless, we also compare the asymptotic errors derived from the Fisher matrix to those determined from the full Monte Carlo sampling of the posterior distribution of $\vlos$ and the stellar atmospheric parameters.

The accuracy of line-of-sight velocity measurement is known to depend on the proper choice of templates and template mismatch introduces additional systematic error \citep{David2014A&A...562A..97D}. In the case of fitting observations with synthetic spectra, template mismatch can mean the following. (i) The template is inaccuare because the synthetic spectra do not match real observations due to shortcomings of the stellar models. (ii) The stellar models are accurate but a template with the wrong atmospheric parameters is used. Although the use of templates derived from inaccuare stellar models to fit $\vlos$ is considered a significant source of systematic errors, we do not investigate this topic in this work. In fact, we fit the simulated observations using the same synthetic spectrum grid that we used to generate the simulated observations. We investigate, however, the magnitude of the bias and systematics caused by mismatched atmospheric parameters and characterize them as the function of signal-to-noise ratio.

We also omit the analysis of sources of additional systematic error terms such as convolving stellar templates with a mismatched line spread function, systematic errors in the wavelength solution due to inaccurate wavelength calibration or time-dependent drifts of the wavelength solution over the course of the night.

Software libraries for direct pixel template fitting use a range of approaches. The approaches vary in the following. (i) Whether fitting is done using individual exposures on a per pixel basis or resampled, stacked spectra. (ii) How the fluxing errors (or template inaccuracies) are corrected for. (iii) Whether errors in the wavelength solution are corrected for or not. (iv) Whether a full Bayesian method is used including priors on the parameters, as opposed to calculating the likelihood function only. (v) Whether only a multivariate maximum finder is used or a full Monte Carlo sampling of the posterior (likelihood) distribution is done to estimate the parameters and their uncertainties. In addition to these, many template-fitting $\vlos$ finders are designed to correct for the peculiarities of a specific instrument that can include fluxing systematics, wavelength shifts from slit miscentering \citep{Sohn2007ApJ...663..960S}, etc.

We developed our of line-of-sight velocity measurement algorithm that also works by fitting stellar templates to observed or simulated spectra on a per pixel basis and supports fitting multiple exposures with potentially different line spread functions. The code allows for a multiplicative flux correction to the observation in the form of a wavelength-dependent polynomial to account for fluxing systematics, as well as imprecisions of the templates. Our approach is very similar to \cite{Koposov2011ApJ...736..146K}, with some additions: We calculate the full covarinace matrix of the parameter errors analytically, also taking the convariances of coefficients of the flux correction polynomial into account, and consider the case of non-uniform priors on $\vlos$ and the template parameters when calculating the full error covariance matrix. Our software implementation is immensely optimized to efficiently perform grid interpolation and convolutions at high resolution with wavelength-dependent kernels. The optimizations were necessary to fit a large number of simulations in a reasonable time.

It is generally useful to be able to estimate the expected uncertainty of a parameter measurement before observations. Scaling relations provide a simple way to approximate $\sigma(\vlos)$ from the spectral resolution, signal-to-noise ratio, bandwidth and spectral type of the target. Based on our empirical results on the uncertainty of the line-of-sight velocity, we propose a modified version of the scaling relation of \cite{HC1992ESOC...40..275H} that also takes the spectral type into account.

The paper is structured as follows. In Section~\ref{sec:former_work}, we review former work related to uncertainty estimation of $\vlos$ and then move over to describing the simulated observation in Section~\ref{sec:simulations}. In Section~\ref{sec:fitting}, we introduce the \textit{significance function}, which provides a convenient mathematical formalism to calculate $\sigma(\vlos)$ analytically, but we relegate the detailed calculation to Appendix~\ref{ax:fisher}\@. We explain our performance optimized template fitting algorithm in Section~\ref{sec:algorithm}, and in Section~\ref{sec:results}, we present the outcome of the simulations. Results are discussed and compared to earlier work in Section~\ref{sec:discussion}, and the paper is concluded in Section~\ref{sec:conclusions}.

\section{Former work}
\label{sec:former_work}

Inferring physical parameters, including line-of-sight velocity from spectra is fundamental and has been in the focus of algorithm development for decades. Although computationally more intensive, direct pixel fitting of model templates have largely replaced Fourier methods, originally to extract the line-of-sight velocity distribution from galaxy spectra \citep[e.g.,][]{RixWhite1992MNRAS.254..389R,Newman2013ApJS..208....5N}, but later more complex and refined spectrum processing pipelines have been developed for single star $\vlos$ measurement as well \citep{Koleva2009A&A...501.1269K, Koposov2011ApJ...736..146K, Cunningham2019ApJ...876..124C, Jenkins2021ApJ...920...92J}.

Among the many other spectrum fitting codes, \cite{Koleva2009A&A...501.1269K} introduced the ULySS software library to fit single stellar spectra as well as to analyze the star-formation history and chemical evolution of composite stellar populations. They emphasized the importance of using a line spread function that closely matches that of the instrument to achieve good results. More recently, \cite{Koposov2011ApJ...736..146K} analyzed VLT/GIRAFFE fiber-fed spectra on a per pixel basis highlighting the advantages of using native detector pixels and likelihood stacking to fit the data. \cite{Koposov2011ApJ...736..146K}, based on earlier work of \cite{Koleva2008MNRAS.385.1998K}, developed a pixel-fitting method that allows for a flux correction to the template in the form of a wavelength-dependent polynomial to account for fluxing systematics in the observations as well as imprecisions of the templates. \cite{Koposov2011ApJ...736..146K} consider the coefficients of the flux correction polynomial nuissance parameters and marginalize them from the joint likelihood of $\vlos$ and the template parameters.

\cite{Koposov2011ApJ...736..146K} and later \cite{Jenkins2021ApJ...920...92J} realized that the sky lines in the raw VLT/GIRAFFE spectra had systematic shifts and developed an algorithm to recalibrate the wavelength solutions of 1D spectra based on the positions of sky emission lines. \cite{Walker2006AJ....131.2114W,Walker2015MNRAS.448.2717W} developed a Bayesian technique to fit spectra of dSph members at medium resolution and found that the posterior probability distribution of $\vlos$ and the atmospheric parameters often become strongly non-Gaussian at low $\SNR$.

To evaluate the effect of template mismatch on measuring $\vlos$, \cite{North2002MNRAS.337.1215N} suggested fitting spectra with several templates and found that template mismatch is not a significant source of systematic errors in case of G, K and M stars. \cite{David2014A&A...562A..97D} quantified the effect of template mismatch as a function of the atmospheric parameters and found deviations of $\Delta \vlos \pm 0.1$~km~s$^{-1}$ in the temperature range $4000 \leq \Teff \leq 6000$~K based on fitting mismatched templates to noiseless spectra. They also showed that the bias can be significantly higher for spectra with relatively broad Balmer lines, Paschen lines or Ca\textsc{ii} triplet. \cite{Walker2015MNRAS.448.2717W} showed -- using Monte Carlo sampling of the posterior distribution of $\vlos$, the atmospheric parameters and the flux correction coefficients -- that $\vlos$ is essentially uncorrelated with the other parameters.

In order to estimate uncertainty of $\vlos$ measurements without detailed simulations, \cite{HC1992ESOC...40..275H} provided a scaling formula to calculate $\sigma(\vlos)$ from signal-to-noise, spectral resolving power and instrument bandwidth and proved the validity of the formula on simulations of high resolution spectroscopic observations. \cite{Bouchy2001A&A...374..733B} derived theoretical constraints on $\sigma(\vlos)$ and emphasized that the spectral type plays an important role in the precision of $\vlos$ measurements. They determined that the local slope of the spectrum -- degraded to the resolving power of the instrument and evaluated at the center of each detector pixel -- determines how much information is carried by the spectrum about Doppler shift.

In the present work, we generalize the formalism proposed by \cite{Kaiser2004}, originally to detect faint point sources in noisy pixelated images. \cite{Kaiser2004} showed how to assign significance to the detections and to calculate unbiased error estimates of the parameters such as the centroid positions. When the method is adopted to spectroscopy of faint objects, the \textit{significance function} becomes independent of the normalization of the observed spectrum and the template, and the location of its maximum will coincide with the maximum of the likelihood function. The formalism allows for calculating the Fisher information matrix at the best-fit parameters analytically and deriving expressions for the uncertainties of the model parameter as well as the Doppler shift, as we will show in Section~\ref{sec:fitting}.

\section{Simulated observations}
\label{sec:simulations}

To assess the uncertainty of Doppler shift estimates from synthetic template fitting, we rely on detailed simulations of spectroscopic observations with known $\vlos$, as well as atmospheric and observational parameters. For this study, we generated 1000 simulated observations, in three different spectrograph arm configurations, of every stellar type listed in Table~\ref{tab:model_params}. The distribution of the observational parameters was the same for all stellar types, representing a realistic dark night similar to the environment on Mauna Kea. Our goal was to sample a broad range of signal-to-noise ratios instead of generating a synthetic catalog.  Hence, magnitudes were sampled uniformly from an interval that will typically be accessible by Subaru PFS \citep{Sugai2014SPIE.9147E..0TS}.  The three spectrograph arm configurations of PFS that we investigated are listed in Table~\ref{tab:instrument}. Simulations were executed for the blue and red arms, with the two different resolution configuration of the latter: low and medium resolution. We denote these with B, R and MR, respectively.  The resolution (FWHM of the line spread function) is approximately constant within each of these arms.  A single pointing of PFS consists of an observation in B plus either R or MR\@.  PFS also has a near-infrared channel, but we do not include it in our simulations.

\begin{deluxetable}{lccc}
    \tablecaption{Assumed parameters of the optical spectrograph arms (B: blue, R: low resolution red, MR: medium resolution red) which are very similar to the parameters of the Subaru PFS instrument.
    \label{tab:instrument}}
    \tablehead{
        & \colhead{B} & \colhead{R} & \colhead{MR}
    }
    \startdata
        $\lambda$ coverage [nm] &
            $380$ - $650$ &
            $630$ - $970$ &
            $710$ - $885$ \\
        pixel dispersion [$\Angstrom$] &
            $0.7$ &
            $0.9$ &
            $0.4$ \\
        $\Delta\lambda$ spectral resolution [$\Angstrom$] &
            $2.1$ &
            $2.7$ &
            $1.6$ \\
        $\Delta v$ velocity resolution [km/s] &
            $130$ &
            $100$ &
            $60$ \\
        $R$ resolving power &
            $2300$ &
            $3000$ &
            $5000$
    \enddata
\end{deluxetable}

\subsection{Simulation software}

Our simulations are based on the Exposure Time Calculator (ETC) code version~5, originally written by \cite{Hirata2012arXiv1204.5151H} and modified by the Subaru PFS team \citep{ETCSubaru} for the fiber-fed PFS instrument. The ETC implements an optical model of the telescope, the wide field corrector and the spectrograph to calculate the line spread function (LSF), the fiber trace width, the effective aperture and vignetting. It also applies the combined transmission function of the atmosphere, the optical elements as well as the quantum efficiency curve of the detector and the model of the detector noise characteristics to calculate the noise terms. To calculate the photon counts and the noise, object photons, sky photon (continuum and lines separately), photons of light scattered by the Moon and stray light from the spectrograph are taken into account.

Since the aforementioned computations can take a significant time, we modified the ETC to write out the line spread function, the transmission function and the contributions of the various noise terms to the final flux variance. Then we ran the ETC on a grid of observational parameters including seeing, the angle of the target and the Moon with respect to the zenith and each other, and the field angle of the target with respect to the optical axis to calculate vignetting. By interpolating these pre-tabulated ETC outputs, simulating a single observation takes only a few hundred milliseconds. In order to generate tens of thousands of simulations, we implemented a parallel spectroscopic simulation library in Python that can handle very fast synthetic spectrum grid interpolation, LSF-convolution and resampling. The highly customizable software suite was originally developed to generate large training sets for machine learning spectrum analysis methods and its details will be published elsewhere.

Our simulations are based on the high resolution PHOENIX library of synthetic stellar spectra \citep{PHOENIX2013A&A...553A...6H}, which are available at a resolving power of $R=500,000$ in the optical.

Whenever signal-to-noise ratios are quoted, we refer to the 95~percentile $\SNR$ per resolution element in the MR arm. If the simulation was performed for the low-resolution R arm, the equivalent $\SNR$ for the MR arm is quoted.

\subsection{Model parameters}
\label{sec:model_params}

\begin{figure}
    \begin{center}
        \includegraphics{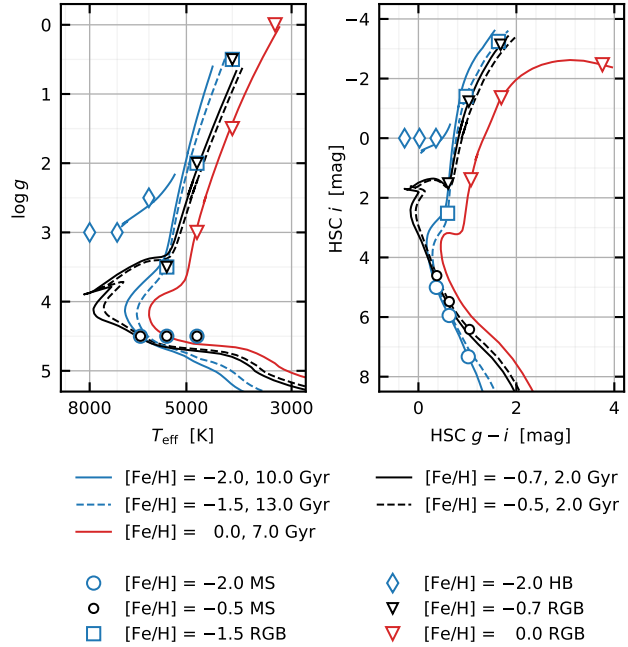}
        \caption{Location of the simulated observations in the plane of atmospheric parameters (left) and  the Subaru Hyper-Suprime~Cam color--magnitude diagram (right). We plot a selection of PARSEC isochrones for reference, as well as the locations of the models we simulated. Note  that the models were chosen to be aligned with the synthetic spectrum grid, thus may not lie on the isochrones.
        See Table~\ref{tab:model_params} for a detailed list of parameters.
        }
        \label{fig:model_params}
    \end{center}
\end{figure}

We selected 18 stellar types that span the range of metallicity, effective temperature and surface gravity that optical Galactic Archaeology surveys typically cover in the Milky Way, dSph galaxies and nearby galaxies with resolvable stars such as M31. Stellar types are summarized in Table~\ref{tab:model_params} and indicated in Figure~\ref{fig:model_params} on top of PARSEC isochrones \citep{PARSEC2012MNRAS.427..127B}, both in terms of the physical parameters and the colors and absolute magnitudes in the Subaru Hyper-Suprime~Cam broadband filter set. The actual parameters slightly differ from the isochrones because the synthetic grid in Table~\ref{tab:model_params} is not exactly aligned with the isochrones. 
For the sake of simplicity, all our models have solar $\aM$ and, when fitting templates to simulations, we fit the three fundamental parameters only.

\begin{deluxetable}{r|c|c|c|c|c}
    \tablecaption{Summary of the stellar models used as the basis for simulated observations. We list the synthetic spectrum atmospheric parameters, stellar evolutionary phase and the number of simulated 15-minute exposures for each model.}
    \label{tab:model_params}
    \tablehead{
        No. & [Fe/H] & $T_\mathrm{eff}$ [K] & $\log g$ & phase & exp. count
    }
    \startdata
    1. & -2.0 & 6250 & 4.5 & MS & 12 \\
2. & -2.0 & 5500 & 4.5 & MS & 12 \\
3. & -2.0 & 4750 & 4.5 & MS & 12 \\
4. & -0.5 & 6250 & 4.5 & MS & 12 \\
5. & -0.5 & 5500 & 4.5 & MS & 12 \\
6. & -0.5 & 4750 & 4.5 & MS & 12 \\
7. & -1.5 & 5500 & 3.5 & RGB & 12 \\
8. & -1.5 & 4750 & 2.0 & RGB & 12 \\
9. & -1.5 & 4000 & 0.5 & RGB & 12 \\
10. & -2.0 & 8000 & 3.0 & HB & 12 \\
11. & -2.0 & 7000 & 3.0 & HB & 12 \\
12. & -2.0 & 6000 & 2.5 & HB & 12 \\
13. & -0.7 & 5500 & 3.5 & RGB & 20 \\
14. & -0.7 & 4750 & 2.0 & RGB & 20 \\
15. & -0.7 & 4000 & 0.5 & RGB & 20 \\
16. & 0.0 & 4750 & 3.0 & RGB & 20 \\
17. & 0.0 & 4000 & 1.5 & RGB & 20 \\
18. & 0.0 & 3250 & 0.0 & RGB & 20 
    \enddata
\end{deluxetable}

\subsection{Simulation details}
\label{sec:simulation_details}

We start the simulations from fluxed synthetic models, which are much higher resolution than the instrument, to be able to evaluate the ability of the template fitting algorithm to detect sub-pixel Doppler shifts. As mentioned in Section~\ref{sec:model_params}, atmospheric parameters were chosen to match the grid so no model interpolation was necessary to generate the simulated spectra. When fitting the simulated observations with stellar templates, and atmospheric parameters are treated as free parameters, we linearly interpolate the grid to calculate the flux of models in between grid points.

We choose the Doppler shift randomly from a uniform distribution and shift the wavelengths of the spectral elements of the high-resolution model before smoothing it with the LSF and interpolating it to the pixels of the detector. The LSF calculated by the ETC already takes the pixelization into account. Hence, we use linear interpolation, since flux-conserving resampling would duplicate the effect of pixelization. We convolve the spectrum, still at high resolution, with the instrumental LSF, expressed as a Gaussian kernel with wavelength-dependent width. The width of the Gaussian is taken from the output of the ETC, and the kernel is evaluated at high resolution, still on the wavelength grid of the input models.

Choosing the Doppler shift for each simulated spectrum randonly from a broad interval, instead of just measuring $\sigma(\vlos)$ at $\vlos = 0$ allows us to test two important things. (i) We can investigate the sensitivity of the template fitting algorithm to sub-pixel shifts and whether there are any systematics depending on how the strong absorption features are shifted with respect to the pixel edges. We ran these tests and found no systematics. (ii) We can better test the part of the template fitting algorithm that provides a coarse initial estimate of $\vlos$ before starting the maximum likelihood optimization.

We assume observational parameters similar to those of at the Mauna Kea Observatories on dark nights, as summarized in Table~\ref{tab:obs_params}. Each of these parameters is sampled uniformly between the given limits. The 15~minute observation time was chosen according to the design parameters of the PFS instrument and we assumed 12~exposures (a total of 3 hours) per field for the MW and dSph targets (model No.~1-12) and 20~exposures (a total of 5 hours) for the M31 targets (model No.~13-18). 
\begin{deluxetable}{lc}
    \tablecaption{Parameters of the simulated spectroscopic observations.
        \label{tab:obs_params}}
    \tablehead{
        parameter & range
    }
    \startdata
        seeing    & $0.6$ - $1.0''$ \\
        extinction $E(B-V)$ & $0.0$ - $0.5$~mag \\
        target zenith angle & $0$ - $60^\circ$ \\
        field angle & $0.0$ - $0.5^\circ$ \\
        single exposure time & $15$~min \\
        exposure count & 12 or 20 \\
        object apparent magnitude & see text \\
        line-of-sight velocity & $-500$ - $500$~km~s$^{-1}$
    \enddata
\end{deluxetable}

Instead of using $\SNR$ as the input parameter, we normalize the synthetic flux of the model spectrum to a magnitude drawn from a uniform distribution that covers the typical magnitude range of potential targets of PFS. Although this method will not result in uniform sampling in $\SNR$, it will make sure that the flux error from sky subtraction, scattered light and the noise contribution of the detector is taken into account at the right ratio. At faint magnitudes, the sky photons are the most significant source of flux error, hence, rescaling the noise level of a simulation to a given $\SNR$ would change the ratio of the noise components. We use the pre-tabulated output from the ETC to obtain the combined transmission curve of the atmosphere and the instrument, as well as the continuum and line emission of the sky. We only simulate dark time observations here but the software is also capable of calculating the scattered light from the Moon. To calculate the resulting signal-to-noise, we approximate the photon noise with a Gaussian distribution and sum up the detector noise terms in quadrature to calculate the variance in each pixel.

We sample the value of the $E(B-V)$ extinction randomly between $0.0$~and~$0.5$~mag and apply the average Milky Way extinction curve of \cite{Cardelli1989ApJ...345..245C}. When fitting the simulated spectra, we do not correct for extinction, but rather let the flux correction polynomial (see Section~\ref{sec:flux_correction}) account for it.

We apply a multiplicative function to the flux in the form of a slowly changing, but high-order function with a maximum amplitude of 2~per~cent. The purpose of this function is to simulate the effects of improper fluxing as well as differences between the continua of synthetic and real spectra.

Even though we do not simulate sky subtraction systematics or residuals, we mask out pixels where the strong sky lines are, including the wings of those lines. We also mask out pixels where the $\SNR$ would be extremely low due to low instrumental transmission.

The actual noise realization is only added to the simulated spectra during template fitting to allow generating multiple noise realizations to imitate multiple exposures. When fitting templates to the simulated observations, we assume 15~min exposures of the target that add up to 3~hr or 5~hr total observation time. We assume the same instrumental noise variance and observational paramaters for each exposure. As explained in Section~\ref{sec:fitting}, the exposures are not stacked, but the templates are fitted on a per pixel basis.

We will also demonstrate how the zero point errors of the wavelength calibration affect the uncertainty of $\vlos$ measurements. To simulate the zero point error, we execute the simulations as explained above, then we apply a small shift to the pixel central wavelengths which corresponds to an affine transformation in $\lambda$ and translation in $\log \lambda$. Simulating the zero point error this way, instead of before the step when we interpolate the spectrum to the detector pixels, is a simple short-cut that makes fitting templates to multiple exposures with different wavelength solutions easier. Note, that the majority of the simulations were generated without wavelength calibration error, hence the results for $\sigma(\vlos)$ represent the theoretical lower limit that does not take the systematic error floor of instrument into account. The instrumental velocity error floor is generally thought to be around or slightly above $1/100^\mathrm{th}$ of a pixel, or approximately $1.3$, $1.0$ and $0.6$~km~s$^{1}$ in the spectrograph arms B, R and MR, respectively.

The steps of the simulation procedure are as follows.
\begin{enumerate}
    \item Load the high-resolution input model from the synthetic spectrum grid. 
    \item Shift the model to the desired Doppler shift between $-500$~and~$500$~km~s$^{-1}$.
    \item Apply the extinction model.
    \item Given a randomly chosen observed magnitude, calculate the flux in physical units. 
    \item Convolve the synthetic spectrum with the instrumental LSF on the original wavelength grid of the high resolution model.
    \item Interpolate the spectrum to the grid of detector pixels.
    \item Calculate the photon count in each pixel from the source, as well as different components of the sky and Moon.
    \item Calculate the signal-to-noise in each pixel.
    \item Apply the model for simulated flux calibration error.
    \item Generate a pixel mask of strong sky lines and very noisy pixels.
    \item Optionally, shift central wavelengths of each pixel to mimic wavelength calibration zero-point error.
\end{enumerate}

\section{Template fitting}
\label{sec:fitting}

In this section, we will follow \cite{Kaiser2004} who introduced a formalism for optimal detection of faint, point-like objects in noisy pixelated images. Here we generalize his approach in the context of spectroscopy, where we consider the optimal extraction of signal from noisy observations by cross-correlating them with a stellar template. While the objective function derived from these calculations coincides with the quantity that spectrum template fitting codes typically aim to maximize, we provide formulae for the full covariance matrix and analytic expression for the uncertainty of $\vlos$.

In the most general case that we consider, targets are observed using multiple spectrograph arms and many detector read-outs. We assume that repeated observations can happen at different times during different atmospheric conditions and with different fiber configurations, where the latter can completely change both the line spread function (LSF) and the pixelization of the spectrum when a different fiber is assigned to the same target. As usual, we do likelihood stacking instead of attempting to resample the spectra to the same wavelength grid and stack the flux. Resampling and stacking, in addition to mixing the LSF from different observations, would result in strong correlations in the flux error. Likelihood stacking, on the other hand, simplifies handling the flux error but has consequences on computational performance that have to be addressed.

\subsection{Analytic calculation of the significance function}
\label{sec:significance}

\cite{Kaiser2004} introduced a formalism to detect faint point sources in images. Here we generalize this formalism for spectroscopy and show how it allows the Fisher information matrix and elements of its inverse characterize the error of $\vlos$ measurements. While \cite{Kaiser2004} assumed a point source convolved with a spatially extended point spread function to cross-correlate with the observed image, we consider a template spectrum convolved with the line spread function and its cross-correlation with the observed spectrum. The template can be fixed or the atmospheric parameters of the template can also be subject to optimization, along with $\vlos$.

While the amplitude of the image of a point source can be described by a single number, in case of spectra, we also have to discuss corrections to incorrect fluxing of the observations or, not quite equivalently, inaccurate model continua. We defer discussing the error arising from inaccurate wavelength calibration and the methods of wavelength correction to a follow up paper. The most generic model we consider here allows for finding $\vlos$, the atmospheric parameters of the template and a multiplicative correction of the observed flux in the form of a linear combination of wavelength dependent basis functions where the coefficients are unknown. When fitting the template parameters, or including correction for the fluxing systematics, calculations quickly become elaborate, hence we moved them to Appendix~\ref{ax:fisher}, and only quote the most important results here.

With $p$ representing the pixel label that indexes each pixel of each exposure, $f_p$ is the sky-subtracted photon count of the observed spectrum, $\sigma_p^2$ is the total noise variance of the observed spectrum. The variance $\sigma_p^2$ is calculated from the total photon count $n_p$, which includes the object photons and sky photons, as well as from additional error terms such as the read-out noise and dark current. The model $m_p(z)$ is the stellar spectrum template, Doppler-shifted by $z=\vlos/c$, convolved with the LSF at high resolution and resampled to the detector pixels. The orthogonal basis functions $q_j(\lambda)$ are defined across the full wavelength coverage and evaluated at the wavelengths of pixel centers $\lambda_p$ to get the matrix $q_{jp} = q_j(\lambda_p)$.

Assuming Gaussian errors of the flux, the likelihood function that includes the template parameters, as well as the flux correction, takes the form of
\begin{equation}
    \begin{split}
    & L(A_j, z, \theta) = \\ 
        & \prod_{p} \frac{1}{\sqrt{2 \pi \sigma_p^2}} 
        \exp \left[ -\frac{1}{2}\frac{\left[ 
            f_p - \sum_j A_j q_{jp} \cdot m_p(z, \theta)
        \right]^2}{\sigma_p^2} \right],
    \end{split}
\end{equation}
where the product goes over the $p$ pixels of each spectrograph arm and each exposure. Evaluated at each pixel center, $m_p$ is the flux of the template which depends on the Doppler shift $z$ and the atmospheric parameters $\theta$, whereas the unknown $A_j$ coefficients form a linear combination with the $q_{jp}$ orthogonal functions. Here we also assume the independence of the flux errors, measured in each pixel. This is a good first approximation if the spectrum is not rebinned, however, the flux error in neighboring pixels is usually covariant due to sky and scattered light subtraction.

If we only consider a single scalar amplitude $A$ instead of a wavelength dependent flux correction, the likelihood becomes
\begin{equation}
    L(A,z) = \prod_{p} \frac{1}{\sqrt{2 \pi \sigma_p^2}} 
        \exp \left[ -\frac{1}{2}\frac{\left[ f_p-A\,m_p(z) \right]^2}{\sigma_p^2} \right]
    \label{eq:likelihood}
\end{equation} 
The log likelihood, with the constant terms omitted, can be written as the sum of two functions in the form of
\begin{equation}
    \mathcal{L}(A,z) = A \varphi(z) - \frac{1}{2} A^2 \chi(z),
    \label{eq:likelihood_phi_chi}
\end{equation}
where
\begin{equation}
    \varphi(z) = \sum_p \frac{f_p m_p(z)}{\sigma_p^2},
    \quad \mathrm{and} \quad
    \chi(z) = \sum_p \frac{m_p^2(z)}{\sigma_p^2}.
    \label{eq:phi_chi}
\end{equation}
In the original problem, \cite{Kaiser2004} considers point sources, so the equivalent of $m_p$ in his case is the point spread function which means $\varphi(z)$ is the double convolution of the ideal image, i.e. the PSF-convolved observed image. In our case, $m_p(z)$ is the spectrum template convolved with the instrumental LSF so $\varphi(z)$ is the cross-correlation function of the observation and the LSF-convolved template, i.e. a matched filter, weighted by the inverse variance. It is interesting to see that $\varphi(z)$ becomes independent of the observation when the object is so bright that the noise is dominated by the object photons and $\sigma_p^2$ becomes approximately equal to $f_p$. When observing faint objects, however, the noise is dominated by the sky photons and $\varphi(z)$ becomes independent of the template. \cite{Kaiser2004} compares $\chi(z)$ to an exposure map whereas in our case $\chi(z)$ can be related to the information content of the spectral pixels.

At the maximum of the likelihood, the partial derivatives of $\mathcal{L}$ should vanish which yields the equations
\begin{equation}
    \frac{\partial \mathcal{L}}{\partial A} = \varphi(z) - A \chi(z) = 0
    \label{eq:max_L_1}
\end{equation}
\begin{equation}
    \frac{\partial \mathcal{L}}{\partial z} = A \varphi'(z) - \frac{1}{2} A^2 \chi'(z) = 0,
    \label{eq:max_L_2}
\end{equation}
where the prime denotes differentiation by $z$. The first equation can be solved for $A$ and gives
\begin{equation}
    A = \frac{\varphi(z)}{\chi(z)},
    \label{eq:amplitude}
\end{equation}
which can, in turn, be substituted into the second one to get
\begin{equation}
    \frac{\varphi'(z)}{\varphi(z)} = \frac{1}{2} \frac{\chi'(z)}{\chi(z)}.
\end{equation}
This latter can be rewritten as
\begin{equation}
    \frac{d}{dz} \left( \frac{\varphi(z)}{\sqrt{\chi(z)}} \right) = 0.
\end{equation}
\cite{Kaiser2004} called
\begin{equation}
    \nu(z) = \frac{\varphi(z)}{\sqrt{\chi(z)}}
    \label{eq:significance}
\end{equation}
the \textit{significance function}. This is the function that needs to be maximized to get the best fit Doppler shift. The significance is only a function of $z$ and the lack of explicit dependence on $A$ expresses the fact that the normalization of the template carries no information on the Doppler shift.

Let us denote the quantities evaluated at the maximum of $\nu(z)$ with the subscript $0$. Since at the maximum of the significance function $\varphi_0=A_0 \chi_0$ holds and $\mathcal{L}_0 = A_0 \varphi_0 - A_0^2 \chi_0$, the maximum of the log-likelihood is at the same $z$ as the maximum of $\nu(z)$ and
\begin{equation}
    \mathcal{L}_0 
        = \frac{\varphi_0^2}{\chi_0} -\frac{1}{2} \frac{\varphi_0^2}{\chi_0} 
        = \frac{1}{2} \nu_0^2.
    \label{eq:likelihood_maximum}
\end{equation}
\subsection{The Fisher information matrix}
\label{sec:fisher_matrix}

We are now at a point to calculate the Fisher information matrix for measuring the line-of-sight velocity from a noisy observation, defined as
\begin{equation}
    F_{Az} = -\Big\langle \frac{\partial^2\mathcal{L} }{\partial A\partial z}\Big\rangle_0,
    \label{eq:fisher}
\end{equation}
where the 0 subscript indicates that this should be evaluated at the parameter values corresponding to the maximum point $z_0$ of the significance function $\nu(z)$, and the associated $A_0$ that can be calculated from Equation~\ref{eq:amplitude}. The averaging in Equation~\ref{eq:fisher} is done over all possible realizations of the noise.

Let us first write down the curvature matrix of a single noise realization which consists of the second derivatives of the-log likelihood:
\begin{equation}
    C = 
    \begin{pmatrix}
        -\dfrac{\partial^2 \mathcal{L}}{\partial A^2} &
        -\dfrac{\partial^2 \mathcal{L}}{\partial A\partial z}\\[\vertical_distance]
        -\dfrac{\partial^2 \mathcal{L}}{\partial A\partial z}& -\dfrac{\partial^2 \mathcal{L}}{\partial z^2}
    \end{pmatrix} =
    \begin{pmatrix}
        -\chi            & -\varphi' +A \chi'\\[\vertical_distance]
        -\varphi'+A\chi' & -A\varphi''+\dfrac{1}{2}A^2\chi''
    \end{pmatrix}. 
\end{equation}
By using Equations~\ref{eq:max_L_1}~and~\ref{eq:max_L_2} at the maximum point, we can simplify this into the Fisher matrix in the form of
\begin{equation}
    F = 
    \begin{pmatrix}
        \chi_0&\varphi'_0\\[7pt]
        \varphi'_0& -\dfrac{\varphi_0\varphi''_0}{\chi_0}
        +\dfrac{1}{2}\varphi^2_0\dfrac{\chi''_0}{\chi^2_0}
    \end{pmatrix},
\end{equation}
in agreement with Equation~22 of \cite{Kaiser2004}. We can calculate the second derivative of the significance $\nu(z)$ as
\begin{equation}
    \dfrac{\nu''}{\nu} = \dfrac{\varphi''}{\varphi}-\dfrac{1}{2}\dfrac{\chi''}{\chi}
        + \dfrac{3}{4}\left(\dfrac{\chi'}{\chi} \right)^2 -\left(\dfrac{\varphi'}{\varphi}\right)
        \left(\dfrac{\chi'}{\chi}\right).
    \label{eq:nu_second_diff}
\end{equation}
At the maximum point this becomes
\begin{equation}
    \dfrac{\nu''_0}{\nu_0} = \dfrac{\varphi''_0}{\varphi_0}-\dfrac{1}{2}\dfrac{\chi''_0}{\chi_0}
        +\left(\dfrac{\varphi'_0}{\varphi_0} \right)^2.
\end{equation}
Multiplying this with $\nu_0^2=\varphi_0^2/\chi_0$, we get
\begin{equation}
    \dfrac{\varphi''_0\varphi_0}{\chi_0}
        -\dfrac{1}{2}\varphi_0^2\dfrac{\chi''_0}{\chi_0^2} = \nu_0 \nu''_0 
        - \dfrac{{\varphi'_0}^2}{\chi_0},
\end{equation}
with which we can rewrite the Fisher matrix in the simpler form of
\begin{equation}
    F = 
    \begin{pmatrix}
        \chi_0&\varphi'_0\\[7pt]
        \varphi'_0& -\nu_0 \nu_0'' +\dfrac{{\varphi'_0}^2}{\chi_0}
    \end{pmatrix}.
    \label{eq:fisher_simple}
\end{equation}
The determinant of $F$ is particularly simple and takes the form of
\begin{equation}
    |F| = -\chi_0 \nu_0 \nu_0''.
\end{equation}
The determinant is always positive as $\nu$ has a maximum so $\nu''$ must be negative. Now we can easily see that the Cram\'er--Rao lower bound of the Doppler shift error is
\begin{equation}
    \sigma_z^2 = -\dfrac{1}{\nu_0\nu''_0}.
    \label{eq:z_error_from_nu}
\end{equation}
Equation~\ref{eq:z_error_from_nu} has no explicit dependence on $A$, the normalization of the template, which means that the lower bound on the uncertainty of the Doppler shift is independent of the normalization (but the errors of $z$ and $A$ are not uncorrelated). The scaling of $\sigma_z^2$ is very intuitive:  $1/\nu_0\nu''_0$ scales with the variance of the flux $\sigma_p$, thus so will $\sigma_z^2$, and the dimensions of $\sigma_z$ are km~s$^{-1}$. In other words, the Doppler shift error is proportional to the inverse of the square root of the typical signal-to-noise ratio of the flux measurements.

\subsection{Including the atmospheric parameters}
\label{sec:significance_model_params}

When fitting synthetic spectrum templates to observations, often not only the Doppler shift, but the atmospheric parameters are also unknown. In this more generic case, in addition to $z$, the template model $m_p(z, \theta)$, consequently $\varphi$ and $\chi$, depend non-linearly on the $\theta$ parameters, and derivatives of the significance function and $\phi$, with respect to $z$ and $\theta$, will appear in the Fisher matrix in the block form of
\begin{equation}
    F = 
    \begin{pmatrix}
        \chi&\varphi_z & \varphi_\beta\\[7pt]
        \varphi_z& -\nu\nu_{zz} +\dfrac{{\varphi_z}^2}{\chi} & -\nu\nu_{z\beta}+\dfrac{\varphi_z\varphi_\beta}{\chi}\\[14pt]
        \varphi_\alpha& -\nu\nu_{\alpha z} +\dfrac{{\varphi_\alpha\varphi_z}}{\chi} & -\nu\nu_{\alpha\beta} +\dfrac{\varphi_\alpha\varphi_\beta}{\chi}
    \end{pmatrix},
    \label{eq:fisher_params}
\end{equation}
where $z$ in the index of $\varphi$ and $\nu$ means partial differentiation by $z$ and the Greek indices indicate differentiation with respect to the $\theta$ template parameters. All functions in Equation~\ref{eq:fisher_params} have to be evaluated at the maximum point of $\nu$. In Appendix~\ref{ax:fisher_model_parameters}, we show that even in this case, the Doppler shift error can be calculated analytically and the result is
\begin{equation}
    \sigma_z^2  = \left(\dfrac{1}{-\nu\nu_{zz}}\right) \dfrac{1}{1 + \sum_\alpha \frac{s_\alpha^2}{\lambda_\alpha}},
    \label{eq:velocity_error_template_params}
\end{equation}
where the sum of always positive numbers in the denominator is related to the total error of the template fit caused by the uncertainty of template parameters. This result is also very intuitive as it expresses that optimizing the template parameters -- in comparison to optimizing for the Doppler shift only with a fixed template -- always decreases the error of the Doppler shift. In general, the more free parameters the stellar template has, the smaller uncertainty the Doppler shift can be estimated with.

\subsection{Linear model for continuum and flux correction}
\label{sec:flux_correction}

Since the flux calibration of observed spectra is often affected by unknown systematics, a wavelength-dependent, non-linear correction function is typically used to correct for fluxing errors, incorrect de-reddening, discrepancies between the templates and real continua, etc. The most practical, generic flux correction function is a low order polynomial in the form of $A(\lambda) = \sum A_n q_n(\lambda)$. The correction function does not depend on the Doppler shift which means that we always correct the observed flux, even when the difference between the templates and the observations comes from the incorrect templates. This is because correcting the flux of the template would require Doppler shifting the correction function, which we want avoid to keep the math simple.

Introducing a wavelength dependent flux correction results in the log-likelihood of
\begin{equation}
    \mathcal{L}(\bA, z) = \bA^T \bphi -\frac{1}{2} \bA^T \! \bchi \bA,
\end{equation}
where we again omitted the additive constants and $\bA$, $\bphi$ and $\bchi$ are now vectors and matrices which take the form
\begin{align}
    \varphi_k(z, \theta) & =
        \sum_p q_{kp} \frac{f_p m_p(z, \theta)} {\sigma_p^2}, \\
    \chi_{kn}(z, \theta) & =
        \sum_p q_{kp} \, q_{np} \, \frac{m_p^2(z, \theta)}{\sigma_p^2},
\end{align}
where $q_{kp}$ are the linearly independent set of vectors introduced earlier.
The linear independence of the $q_{kp}$ vectors will ensure that the matrix $\chi$ is invertible and $\bA$ can be calculated as
\begin{equation}
    \bA = \bchi^{-1}\bphi.
\end{equation}
We can now write down the curvature matrix in the block form of
\begin{equation}
    \begin{split}
    & C = \\
        & \begin{pmatrix}
                - \bchi &
                    \bphi_z - \bchi_z \bA & 
                        \bphi_\beta - \bchi_\beta \bA \\
                \bphi^\intercal_z - \bA^\intercal \bchi_{z} &
                    \bA^\intercal \! \bphi_{zz} - \frac{1}{2} \bA^\intercal \! \bchi_{zz} \bA &
                        \bA^\intercal \! \bphi_{z\beta} - \frac{1}{2} \bA^\intercal \! \bchi_{z\beta} \bA \\
                \bphi^\intercal_\alpha - \bA^\intercal \bchi_\alpha &
                    \bA^\intercal \! \bphi_{\alpha z} - \frac{1}{2} \bA^\intercal \! \bchi_{z\alpha} \bA &
                        \bA^\intercal \! \bphi_{\alpha\beta} - \frac{1}{2} \bA^\intercal \! \bchi_{\alpha\beta} \bA
        \end{pmatrix},
    \end{split}
\end{equation}
where $z$ in the index of $\bphi$ and $\bchi$ means element-wise differentiation with respect to the Doppler shift and the Greek letters in the index mean differentiation by each of the template parameters $\theta$.

The corresponding significance function depends only on the Doppler shift and the template parameters and can be defined as
\begin{equation}
    \frac{\nu^2(z, \theta)}{2} = \bphi^T \! \bchi^{-1} \bphi.
    \label{eq:significance_fluxcorr}
\end{equation}
This is still a scalar function with no explicit dependence on the flux correction coefficients. It has the maximum at the same values of $z$ and $\theta$ where the likelihood function does. We can now write down the Fisher matrix in a block matrix form as
\begin{equation}
    F = 
        \begin{pmatrix}
                \bchi &
                    - \bphi_z& 
                        - \bphi_\beta \\
                - \bphi^T_z &
                    - \nu\nu_{zz} + \bphi^\intercal_z \! \bchi^{-1} \bphi_z &
                        - \nu\nu_{z\beta} + \bphi^\intercal_z \! \bchi^{-1} \bphi_\beta \\
                -\bphi^T_\alpha &
                    - \nu\nu_{\alpha z} + \bphi^\intercal_\alpha \! \bchi^{-1} \bphi_z &
                        - \nu\nu_{\alpha\beta} + \bphi^\intercal_\alpha \! \bchi^{-1} \bphi_\beta \\
        \end{pmatrix},   
    \label{eq:fisher_full} 
\end{equation}
where all expressions have to be evaluated at the parameters that maximize the significance function.
According to a calculation similar to the previous case described in Section~\ref{sec:significance_model_params}, the uncertainty of the Doppler shift will be the same as in Equation~\ref{eq:velocity_error_template_params}.

The Fisher matrix in Equation~\ref{eq:fisher_full} can be evaluated numerically by calculating the Hessian of the significance function around the maximum, as well as the element-wise first partial derivatives of the vector $\bphi$. We must point out that the numerical differentiation happens with respect to $z$ and the template parameters only and not with respect to the flux correction coefficients. This is advantageous, since algorithms for numerical evaluation of the Hessian typically scale as $\mathcal{O}(D^2)$, where $D$ is the number of parameters, but adaptive algorithms that attempt to reduce the errors of the differentials can scale even worse. Calculating the Fisher matrix using our formalism efficiently reduces the number of necessary function evaluations as there is no differentiation with respect to the flux correction coefficients. Yet, the inverse of Equation~\ref{eq:fisher_full} yields the errors and covariances of the flux correction coefficients, as well as the template parameters $\theta$ and $z$. When twice differentiating the significance functions with respect to $\vlos$ and the three fundamental template parameters, computation of the Hessian still requires several hundred function evaluations using the \texttt{numdifftools} Python package.

\subsection{Bayesian formalism with priors}

When fitting stellar spectra with templates with unknown parameters, one can often benefit from \textit{a priori} knowledge about the spectral type of the observed object. For example, photometric data can constrain the effective temperature and provide some limits on surface gravity. To incorporate this into a probabilistic model, most often the Bayesian posterior probability is considered which, in the generic case of optimizing for the template parameters and the flux correction coefficients, can be written as
\begin{equation}
    p(\bA, z, \theta \given D) =
        \frac{L(\bA, z, \theta) p(\bA) p(z) p(\theta)}
             {\int \! L(\bA, z, \theta) p(\bA) p(z) p(\theta) \dd \bA \dd z \dd \theta},
    \label{eq:posterior}
\end{equation}
where $D$ is a symbol for the observations and $p(z)$ and $p(\theta)$ are the prior probability distributions defined on the Doppler shift and the atmospheric parameters. We will soon see that the prior $p(\bA)$ on the flux correction coefficients must be flat in order to follow the same formalism as in Section~\ref{sec:flux_correction}. The integral in the denominator of Equation~\ref{eq:posterior} is the usual normalization constant that appears in Bayesian posteriors. It is not necessary to calculate it, as the location of the maximum of the posterior does not depend on it, and Monte Carlo sampling of $p(z, \theta \given D)$ requires a function that is only proportional to the posterior.

The location of the maximum of Equation~\ref{eq:posterior} coincides with the location of the maximum of its logarithm, hence, after dropping the additive constants that come from the denominator or elsewhere, the log-posterior will be the sum of the log-likelihood and the logarithm of the priors as
\begin{equation}
    \mathcal{P}(\bA, z, \theta \given D) = \mathcal{L}(\bA, z, \theta) + \pi(z) + \pi(\theta),
    \label{eq:log_posterior}
\end{equation}
where we denoted the logarithm of the priors with $\pi(z)$ and $\pi(\theta)$ and already assumed that $p(\bA)$ is flat, hence its partial derivatives are zero. The location of the maximum of Equation~\ref{eq:log_posterior} can be found by equating its partial derivatives to zero, as before, in the case of the maximum likelihood method. Since $\pi(\bA) = 0$, we can write down the same system of equations for $A$ as in Equations~\ref{eq:max_L_1}~and~\ref{eq:max_L_2} but $\varphi$ and $\chi$, as well as the significance function $\nu$, as defined in Equation~\ref{eq:significance_fluxcorr}, will also depend on the $\theta$ model parameters. Similarly to Equation~\ref{eq:likelihood_maximum}, at the optimum of the amplitude, the log-posterior can be rewritten as
\begin{equation}
    \mathcal{P}(\bA_0, z, \theta \given D) = \frac{1}{2} \nu^2(z, \theta) + \pi(z) + \pi(\theta),
    \label{eq:log_posterior_significance}
\end{equation}
where the 0 index expresses that the posterior and the significance function are to be taken at the optimum with respect to the flux correction coefficinets $\bA$. The Doppler shift and template parameters maximizing the posterior can be found using the same non-linear maximization methods as in the case of Equation~\ref{eq:likelihood_maximum}.
When the priors on $z$ and each of the $\theta$ parameters are independent, no mixed second derivatives of them will appear in the Fisher matrix, which takes the form of
\begin{equation}
    \begin{split}
    & F = \\
        & \begin{pmatrix}
                \bchi &
                    - \bphi_z& 
                        - \bphi_\beta \\
                - \bphi^T_z &
                    - \nu\nu_{zz} + \bphi^\intercal_z \! \bchi^{-1} \bphi_z - \pi_{zz} &
                        - \nu\nu_{z\beta} + \bphi^\intercal_z \! \bchi^{-1} \bphi_\beta \\
                -\bphi^T_\alpha &
                    \!\!\!\! - \nu\nu_{\alpha z} + \bphi^\intercal_\alpha \! \bchi^{-1} \bphi_z &
                        \!\!\!\! - \nu\nu_{\alpha\beta} + \bphi^\intercal_\alpha \! \bchi^{-1} \bphi_\beta - \pi_{\alpha\beta} \\
        \end{pmatrix}.
    \end{split}
    \label{eq:fisher_prior} 
\end{equation}

In practice, we find that the variances of the flux correction coefficients are several orders of magnitude smaller than the variances of the Doppler shift and the template parameters. The same is true for the covariances between the flux correction coefficients and the template parameters, including the Doppler shift. For example, \cite{Walker2015MNRAS.448.2717W} solved a stellar template fitting problem by Monte Carlo sampling the posterior of a hierarchical Bayesian model very similar to Equation~\ref{eq:posterior}. They treated the flux correction coefficients as random variables and found no correlations between the variances of the coefficients and the rest of the model parameters including $z$ and the atmospheric parameters. It is general practice, therefore, to determine the covariance matrix of the uncertainty of the line-of-sight velocity and the template parameters by taking the flux correction coefficients at the location of the maximum significance and calculating the covariance matrix from the inverse of the Hessian of the likelihood function with respect to $z$ and $\theta$ only.

\section{Algorithms}
\label{sec:algorithm}

We have developed our own spectral template-fitting software library that integrates with the observation simulation code and implements the significance maximization formalism outlined in Section~\ref{sec:significance}, and, in more detail, in Appendix~\ref{ax:fluxcorr}.

Our template fitting algorithm is straightforward but we have two important goals: precision and performance. In order to fit $\vlos$ with sub-pixel precision, we decided to use as high resolution templates as possible and perform the LSF convolution at the original resolution of the synthetic spectra. High performance implementations of both the simulation code and the fitting were necessary to be able to execute the large number of simulations and fit the simulated spectra in many configurations, including full Monte Carlo sampling of the posterior distribution. In particular, we focused on high performance convolution and synthetic spectrum grid interpolation.

To evaluate the significance function, the following procedure is followed.
\begin{enumerate}
    \item The synthetic spectrum grid is interpolated to the given values of the atmospheric parameters.
    \item The interpolated template is shifted to the given value of $z$.
    \item The template is convolved with the LSF.
    \item The template is interpolated to the pixels of the observed spectrum, for each and each exposure.
    \item The quantities $\bchi$, $\bphi$, and the logarithm of the priors are evaluated.
    \item The value of $\bchi^{-1}\phi$ is calculated.
\end{enumerate}

Step~3 of the algorithm can be performed in conjunction with Step~1 to optimize perfomance, as we will explain in the next section.

\subsection{Convolution with the line spread function}

When simulating spectra, as well as when processing the templates while fitting the simulated observations, we convolve the synthetic spectra with the line spread function at the original resolution of the synthetic spectrum grid, followed by resampling to the detector pixels. It is important to point out that the line spread function we use already accounts for the pixelization, consequently, resampling consists of a simple linear interpolation of the LSF-convolved high resolution template to the center of the pixels instead of flux-conserving rebinning, which would take pixelization into account twice. As a consequence, convolution and interpolation are not interchangeable and the convolution must be done first, in the high resolution representation.

The convolution of templates with a wavelength-dependent line spread function is implemented using a data compression method based on Principal Component Analysis (PCA). This algorithm is applicable when the template spectrum uses regular binning, either linear or logarithmic. To calculate the convolution of a pixelated spectrum $f_p$ with a kernel $k(\lambda)$ that has a non-linear dependence on wavelength, one wants to evaluate
\begin{equation}
    \hat{f}_p =
        \left[ f \star k(\lambda) \right]_p =
        \sum_{n=-N}^{N} f_p \left[ k(\lambda_p) \right]_{p - n},
    \label{eq:lsf_convolution}
\end{equation}
where $\left[ k(\lambda_p) \right]_i$ is the kernel evaluated at the wavelength $\lambda_p$ of pixel $p$ and discretized on the pixel grid around the central pixel. The value $N$ is the half kernel size necessary to evaluate the convolution with some prescribed precision, i.e. the kernel either has a finite support or its value is negligible outside the $[-N; N]$ interval. The kernel $k(\lambda)$ is usually a high order function and calculating the convolution requires at least $P \times (2 N + 1)$ function evaluations, where $P$ is the number of pixels. As a consequence, when convolving many spectra with the same kernel, pre-tabulating the kernel around each pixel is an obvious way of optimization. We denote the pre-tabulated kernel by $\left[ k(\lambda_p) \right]_i$, around pixel $p$ where the $i$ index runs over the neighboring pixels. A downside of pre-tabulated kernels, especially when the synthetic spectra are high resolution, is that convolution algorithms require shuffling around very large arrays in memory. Here we show how to leverage the linearity of the convolution operation to reduce the variable-kernel convolution into the sum of a few fixed-kernel convolutions which has very fast implementations available.

Let us evaluate $\left[ k(\lambda_p) \right]_i$ for each pixel $p$ and shift $i \in [-N; N]$ and express the variable kernel as a linear combination of some \textit{basis kernels} $K_{pi}$ in the form of
\begin{equation}
    K_{pi} = \left[ k(\lambda_p) \right]_i = \sum_{j} a_{p\!j} E_{ji},
\end{equation}
where $a_{p\!j}$ are wavelength dependent coefficients and $E_{ji}$ is a matrix consisting of some orthogonal basis vectors. To find the most optimal basis, we apply PCA to $K_{pi}$ and factorize its product with its transpose using singular value decomposition, which yields
\begin{equation}
    K_{i\!p} K^\intercal_{p\!j} = \sum_{l} E_{il} \Sigma^2_l E_{l\!j},
\end{equation}
where the $E_{il}$ matrix consists of the eigenvectors of ${\mathbf{KK}}^\intercal$ and $\mathbf{\Sigma}$ is a diagonal matrix of its eigenvalues. We can then determine the $T_{pi}$ \textit{principal components} of the kernel at each pixel as $T_{pi} = \sum_j K_{pj} E_{ji}$ and reconstruct the kernel from its principal components as 
\begin{equation}
    K_{pi} = \sum_j T_{p\!j} E_{ji}.
    \label{eq:pca}
\end{equation}
The sum over $j$ in Equation~\ref{eq:pca} goes over $2 N + 1$ values, the width of the tabulated kernel. Data compression in PCA is achieved by limiting the sum in Equation~\ref{eq:pca} to only a few eigenvectors belonging to the largest eigenvalues of $\mathbf{K}\mathbf{K}^\intercal$. This data compression technique is lossy, but in practice, varying width Gaussian kernels can be compressed into the first $5$-$10$ principal components with an accuracy of $10^{-5}$. Compare this to the typical kernel width of $N > 50$ when convolving down the high resultion templates to the low resolution PFS instrument.

Most importantly, the first few $T_{p\!j}$ principal components can be interpolated to any intermediate wavelength or fitted with low degree polynomials as a function of wavelength and evaluated at any wavelength. Hence, the kernel can be reconstructed with high accuracy at any wavelength -- as long as the spectral bins remain regular.

To evaluate the convolution of Equation~\ref{eq:lsf_convolution} in this compressed representation, one has to compute
\begin{equation}
    \hat{f}_p =
        \sum_j \sum_{n=-N}^{N} T_{p\!j} f_p E_{n\!j}.
    \label{eq:lsf_convolution_pca}
\end{equation}
When the kernel is compressed with PCA, the sum over $j$ in Equation~\ref{eq:lsf_convolution_pca} goes over only a few values, hence we replaced the expensive array reshuffling necessary to evaluate the convolution using a pre-tabulated kernel with a few fast, optimized fixed-kernel convolutions and a summation.

\subsection{Synthetic grid interpolation combined with \\ convolution and caching}

We chose our input models of the simulations to have atmospheric parameters that are on the grid points of the synthetic spectrum grid, but when fitting the simulated observations with free template parameters, interpolation of the synthetic stellar grid is inevitable. At the same time, when working with high resolution spectra, one needs to make flux interpolation as fast as possible. By ruling out higher order methods for performance reasons, we opted for a simple linear interpolator which only takes the $2^D$ spectra in the corners of a grid cell into account, where $D$ is the number of grid parameters.

The simple linear interpolations of the synthetic spectrum grid offers several optimization opportunities, especially when the convolution step can be combined with the interpolation step. One has to recognize, that iterative optimization algorithms, Monte Carlo samplers and numerical Hessian computation methods take very small steps in the direction of each parameter around the most optimal parameters. Consequently, once convergence has been reached, interpolation will only happen within the very same grid cell. (Unless, of course, the best fit parameters fall onto a grid cell boundary.) Pre-computing and caching the LSF-convolved synthetic spectra at the grid points surrounding the cell, and interpolating the convolved templates can save a significant amount of processing. On the other hand, convolution with the LSF is normally done on the template only \textit{after} it is shifted to some non-zero, but typically small line-of-sight velocity during template fitting. Caching and reusing the LSF-convolved templates for interpolation is only possible when the wavelength dependence of the LSF is weak compared to the wavelength shifts and the order of the convolution and wavelength shift operations can be exchanged. 

\subsection{Maximizing the significance function}

We find the maximum of the significance function using the classical Nelder--Mead method. When applying a polynomial flux correction to the observed spectra, one has to evaluate the significance function according to Equation~\ref{eq:significance_fluxcorr} which involves a matrix inversion. Instead of computing the inverse matrix $\bchi^{-1}$, it is numerically much more stable to solve the equation $\bchi \mathbf{x} = \bphi$ for the vector $\mathbf{x}$ which will yield $\mathbf{x} = \bchi^{-1}\bphi$. Numerical stability of this operation also dictates that the flux values of the template spectrum have to be scaled by a factor to match the interval of typical flux values of the observation.

When template fitting is done in a Bayesian setting and priors are defined on the template parameters, the posterior probability distribution of the parameters is given by Equation~\ref{eq:log_posterior_significance}. The maximum aposteriori solution can be found the same way as the maximum significance solution.

\subsection{Uncertainty estimators of $\vlos$}

In Section~\ref{sec:fisher_matrix}, we derived an analytic formula for the uncertainty of the line-of-sight velocity. The error estimator based on the Fisher matrix, also called the asymptotic error, gives only a lower bound on $\sigma(\vlos)$ and this limit is only achievable in practice when the likelihood function is strictly Gaussian about the maximum and has no local maxima in the few-$\sigma$ vicinity of the global maximum. This is rarely the case when dealing with low signal-to-noise spectra, as noise ``smears'' the likelihood function so that its maximum will be smaller and the curvature at the maximum larger, increasing the influence of nearby local maxima. Hence, precise characterization of $\sigma(\vlos)$ in low signal-to-noise situations requires Monte Carlo sampling of the significance function. We are going to demonstrate this in Section~\ref{sec:results}.

In addition to the asymptotic uncertainty estimator derived from the Fisher matrix, we also determine the full covariance matrix of the parameters from full Monte Carlo sampling of the Bayesian posterior probability distribution when $\vlos$, as well as the template parameters are treated as unknown.

When we present our results in Section~\ref{sec:results}, we are going to compare the uncertainty estimates from the Fisher matrix and MC sampling to those that we measure from an ensemble of repeated simulations.

\subsection{Monte Carlo sampling of the posterior}

We implemented an adaptive Monte Carlo algorithm to generate samples of the significance function or the Bayesian posterior probability distribution. The sampler works by alternating between $\vlos$ and the rest of the parameters. In case of high $\SNR$ spectra when $\vlos$ is well constrained, this Gibbs-like sampling resulted in better mixing of the MC chains than sampling all parameters together. The proposal distributions of both semi-steps are initialized from the covariance matrix of the parameters, as determined from the maximum aposteriori template fitting. The initial states were generated from the maximum aposteriori best fit parameters by perturbing them with a random vector drawn from a multivariate Gaussian distribution with the same covariance matrix as the step proposal. Initializing the state and the step proposal distributions this way allowed us to minimize the length of the burn-in phase and reach a high acceptance rate sooner.

\section{Results}
\label{sec:results}

We generated 1000 realizations of the simulated observed spectra for each stellar type listed in Table~\ref{tab:model_params} and fitted them using different combinations of the spectrograph arms and template spectra, as well as various different configurations of the template fitting algorithm. In the case of maximum likelihood and maximum a posteriori fitting, we calculated $\Delta \vlos$ as the difference between the input line-of-sight velocity and the best velocity estimate obtained by fitting the templates to the simulations. 

The results we present here do not account for some instrumental systematics, such as the wavelength calibration. As a result, velocity uncertainties below the typical value equivalent to $1/100$ of an detector pixel (approximately $0.6$~km~s$^{-1}$ in the MR arm) are unrealistic. Nevertheless, we plot these results as they establish the theoretical lower limit on the uncertainty of $\vlos$. The instrumental error floor is considered to be due to systematics and wavelength calibration errors, neither of which were taken into account when we ran the simulations. We are going to discuss this issue in some detail in Section~\ref{sec:error_estimators}.

When simulating the observations, we took the typical fluxing errors into account by multiplying the flux with a random, slowly changing function that alters the flux by about 2\%, see Section~\ref{sec:simulations}. In addition to this, we applied the reddening law of \cite{Cardelli1989ApJ...345..245C} with a randomly chosen value of $A_V$ between $0$ and $0.5$ and but we do not explicitly correct for this reddening when fitting the templates. Instead, we use a fifth-order Chebyshev polynomial over the entire wavelength range to correct both the fluxing error and reddening, as described in Section~\ref{sec:fitting}.

We fitted the simulated spectra with templates using various configurations. In the most optimistic scenario, the atmospheric parameters of the template were chosen to match the input parameters. In case the atmospheric parameters of the template are also optimized for, instead of solving the maximum significance problem, we performed a full Bayesian maximum a posteriori fit to the simulations. The priors used in the Bayesian model are normally distributed in a finite interval centered on the input parameters, as summarized in Table~\ref{tab:priors}.

\begin{deluxetable}{lllll}
    \tablecaption{Parameters of the priors used to fit the simulations when the template parameters are treated as unknowns. The probability distributions are always centered on the input parameters of the simulations and the limits are calculated with respect to the input parameter values.
                  \label{tab:priors}}
    \tablehead{
        parameter & distribution & parameters & limits & unit
    }
    \startdata
        [Fe/H] & truncated normal & $\sigma = 0.5$ & $\pm 1.0$ & dex \\
        $\Teff$ & truncated normal & $\sigma = 50$ & $\pm 250$ & K \\
        $\logg$ & truncated normal & $\sigma = 1.0$ & $\pm 2.0$ & \\
    \enddata
\end{deluxetable}

\subsection{Signal-to-noise}

Throughout this section, we adopt the following definition of signal-to-noise, denoted with $\SNRp$, when talking about the \textit{per pixel} signal-to-noise of a simulated spectrum. The signal is taken as the flux in a pixel of the noiseless spectrum, convolved down to the resolution of the instrument and resampled to the pixels of the detector. The noise is taken as the standard deviation of the Gaussian noise calculated from the observational parameters, the sky model and the instrumental model, as described in Section~\ref{sec:simulations}. 

To characterize the signal-to-noise ratio of an entire spectrum by a single $\SNR$ number, we take the 95~percentile of the per pixel $\SNRp$ values. The advantage of using a high percentile to calculate the overall $\SNR$ as opposed to the mean or the median is that the 95~percentile is less sensitive to the noisy pixels of the deep absorption lines and more sensitive to the continuum pixels. 

\begin{figure}
    \begin{center}
        \includegraphics{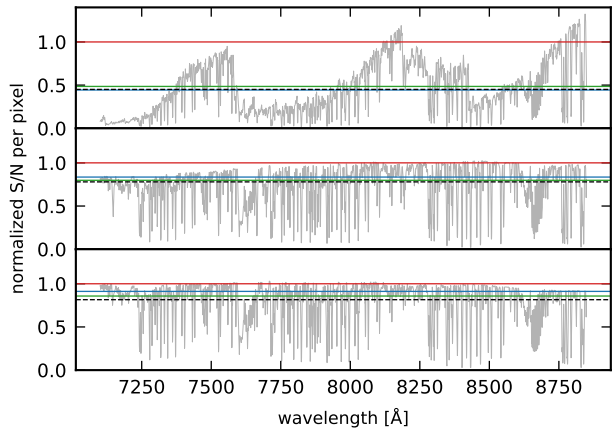}
        \caption{Signal-to-noise per pixel for three different stellar types (from top to bottom: solar metallicity M giant, metal-poor M dwarf and metal-poor K giant). The 95~percentile $\SNR$ is plotted in red and the median $\SNR$ in blue, whereas the dashed black line indicates the $\SNR$ as defined by \protect\cite{Stoehr2008ASPC..394..505S}. See the text for discussion.}
        \label{fig:snr_methods}
    \end{center}
\end{figure}

Characterizing a spectrum with a single $\SNR$ value has its caveats, however. In Figure~\ref{fig:snr_methods}, we plot $\SNRp$ for three stellar spectra that show different levels of absorption in the medium resolution red arm. In the plots, $\SNRp$ is normalized to have a 95~percentile of $1$~per~pixel, indicated by the red line, and the median of $\SNRp$ is plotted in blue. While the 95~percentile $\SNR$ traces the highest signal-to-noise pixels of continuum remarkably well in case of the higher temperature models, it is no longer a good measure when there is no well defined continuum, as in the case of the cool M~giant in the top panel of Figure~\ref{fig:snr_methods}. On the other hand, using the maximum of $\SNRp$ in place of the 95~percentile to characterize a spectrum would not be robust enough.

\cite{Stoehr2008ASPC..394..505S} suggested measuring signal-to-noise from a single realization of a noisy spectrum in the form of
\begin{equation}
    \SNR = \frac{\sqrt{6}}{1.482602} \cdot
           \frac{\mu(f_p)}
                {\mu(\left| 2 f_p - f_{p-2} - f_{p+2} \right|)},
\end{equation}
where $\mu$ denotes the median and $f_p$ is the flux in pixel $p$. This is a robust way of estimating the signal-to-noise from spectra when the signal and the variance are not known individually, only the noisy flux. For reference, we plot this quantity in Figure~\ref{fig:snr_methods} as well, noting that it is subject to the same dependence on spectral type as the mean or the median, hence cannot characterize well the typical $\SNR$ of continuum pixels.

For easy comparison, we will refer to the 95~percentile $\SNR$ as measured in the medium resolution red arm, even when plotting the results for different spectrograph arms, or a combination of arms. Whenever the signal-to-noise is expressed \textit{per resolution element}, as opposed to \textit{per pixel}, we calculate it by multiplying the 95~percentile value of per pixel signal-to-noise by the square root of the typical number of pixels per resolution element. Based on Tab.~\ref{tab:instrument}, the multipliers are $\sqrt{3}$, $\sqrt{3}$ and $\sqrt{4}$ for the B, R and MR arms, respectively.

\subsection{Empirical error of $\vlos$ from repeated simulations}

\begin{figure*}
    \begin{center}
        \includegraphics{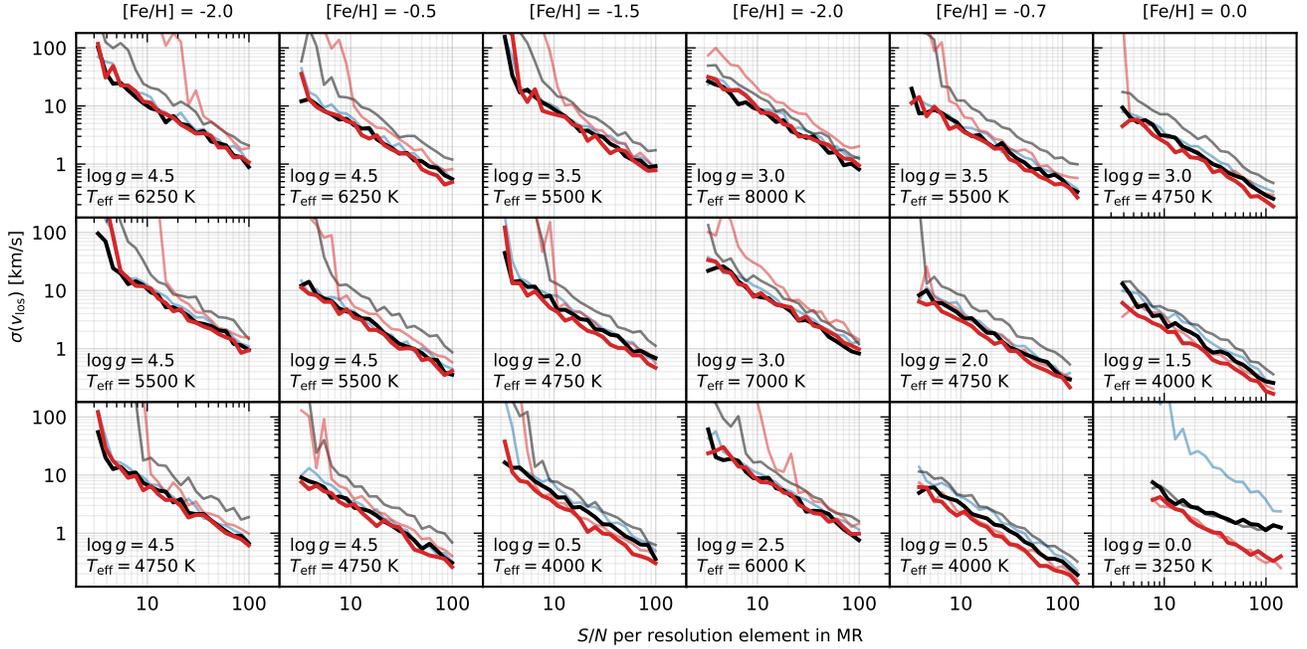}
        \caption{Comparison of the typical empirical line-of-sight velocity error, without any instrumental systematics, for the models listed in Tab.~\ref{tab:model_params} using the Subaru~PFS blue (B -- thin blue curves), red (R -- thin red curves) and medium resolution red (MR -- thin gray curves) arms, as well as the combination of B+R (thick black curves) and B+MR (thick red curves). The curves show the standard deviation of the line-of-sight velocity error $\Delta \vlos$, with respect to the model input, in logarithmic bins of $\SNR$. For purposes of easy comparison, the values of $\SNR$ on the horizontal axes are always the 95~percentile signal-to-noise in the medium resolution red arm only, even when the fitting was done using different arms or a combination of arms.}
        \label{fig:rv_error}
    \end{center}
\end{figure*}

Figure~\ref{fig:rv_error} shows the primary result of this work: the standard deviation of $\Delta \vlos$, calculated in logarithmic bins of $\SNR$ per resolution element. $\SNR$ is quoted for the single arm MR configuration, even when the curves correspond to the R arm or a combination of two arms. We will refer to the standard deviation of $\Delta \vlos$ as the \textit{empirical error} and simply denote with $\sigma(\vlos)$. We determined the empirical errors for fitting each spectrograph arm separately (thin curves) and in the blue and red arms in combination (thick curves). The colors represent the arms used for fitting: B -- blue arm (blue curves), R -- red arm in low resolution configuration (red curves) and MR -- red arm in medium resolution configuration (black curves). We verified that the slope and y-intercept of these curves are insensitive to the binning in $\SNR$: using logarithmic bins in $\SNR$ yields very similar curves regardless of the number of bins, \textit{c.f.}~Figure~\ref{fig:rv_err_bouchy_hc}.

The curves in Figure~\ref{fig:rv_error} show the general trend of $\sigma(\vlos)$ decreasing with the square root of $\SNR$, which is expected based on what we calculated in Section~\ref{sec:fisher_matrix} from the Fisher matrix. Deviations are visible from this trend at low $\SNR$ in most cases and at high $\SNR$ in case of the M~giant. We discuss these deviations in Section~\ref{sec:error_estimators}.

\subsection{Error estimators of $\vlos$ for single spectra}
\label{sec:error_estimators}

In the previous section, we looked at the empirical error of $\vlos$ as a function of $\SNR$ which was calculated from a large number of simulations as the standard deviation of $\Delta \vlos$ in logarithmic bins of $\SNR$. Let us now compare the empirical error to uncertainty estimators that can be determined for a single simulation or a non-repeated observation, such as the one we calculated in Section~\ref{sec:fisher_matrix} from the Fisher information matrix or from Monte Carlo sampling of the likelihood function or the Bayesian posterior.

\begin{figure}
    \begin{center}
        \includegraphics{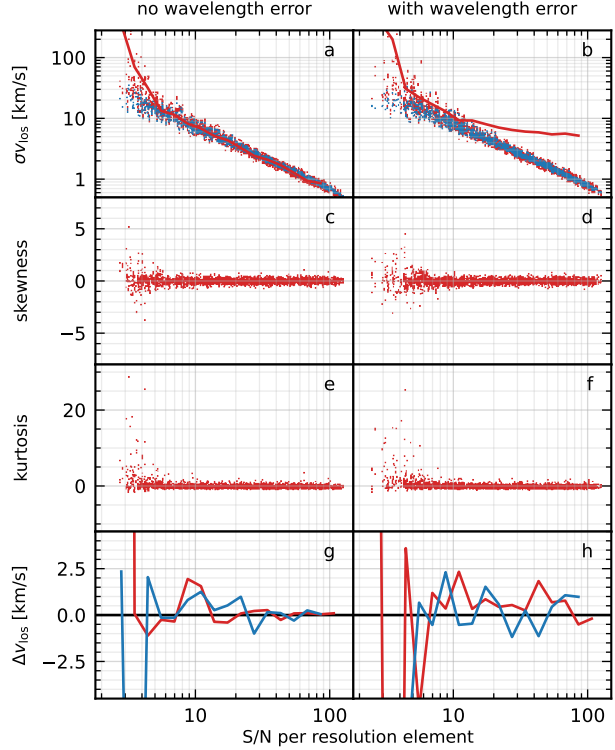}
        \caption{Comparison of error estimates and bias for a model with atmospheric parameters $\MH = -1.5$, $\Teff = 5500$~K and $\logg = 3.5$ \sout{`dSph G subgiant'}, for the  case of fitting $\vlos$ plus the atmospheric parameters, both with and without simulated wavelength calibration error when fitting of arms B+MR in combination. In panels (a) and (b), the blue dots indicate the asymptotic error calculated from the Fisher matrix at the maximum value of the significance function, red dots show the standard deviation of Monte Carlo samples drawn from the Bayesian posterior, and the red curve shows the empirical error calculated from an ensemble of simulations, all in logarithmic bins of $\SNR$. For this particular stellar type, different error estimates start to diverge below $\SNR < 6$ and the error estimator from the Fisher matrix is no longer a good measure of the uncertainty. Panels (c)-(f) show the skewness and the kurtosis of the Monte Carlo samples drawn from the Bayesian posterior, to demonstrate that the posterior marginalized for the atmospheric parameters is no longer Gaussian at low $\SNR$. In panels (g) and (h), the red curve indicates the expectation value of the bias $\Delta \vlos$, as determined from a Monte Carlo sample of the likelihood function, while the blue curve shows the same bias determined from the maximum aposteriori method. Both bias curves are calculated in logarithmic bins of $\SNR$. The bias shows no significant trend over this signal-to-noise range, but its large scatter at low $\SNR$ is obvious. These plots are qualitatively the same for lower temperature, but more luminous, metal-poor giant stars.}
        \label{fig:rv_err_methods}
    \end{center}
\end{figure}

To compare the various estimators of uncertainty, we fitted simulated spectra of a low metallicity G subgiant, typical of stars found in dwarf spheroidal galaxies. We performed the fitting in two configurations, in order to test the effect of an uncertain wavelength calibration. First, we assumed perfectly matching wavelength solutions for each of the 12 exposures in both the B and MR arms. Second, we kept the flux in each pixel, but multiplied the wavelength values of each exposure with a random Doppler shift sampled from a normal distribution with a standard deviation of $25$~km~s$^{-1}$. Since the wavelength errors simulated this way are independent random variables, they are expected to average out according to $\sqrt{N_\mathrm{exp} \cdot N_\mathrm{arm}}$. Consequently, the amplitude of the error was chosen to be very large to emphasize the effect. In both cases, with and without wavelength errors, we optimized for the three atmospheric parameters, and a fifth-order Chebyshev polynomial was used to correct the simulated systematics of the fluxing. The template was fitted using a maximum finder algorithm, as well as Monte Carlo sampling of the posterior probability distribution.

In panels (a) and (b) of Figure~\ref{fig:rv_err_methods}, we plot the empirical error (solid red line) calculated from an ensemble of simulations, in logarithmic bins of $\SNR$, the asymptotic error derived from the Fisher matrix (blue dots) for each simulation and the standard deviation of Monte Carlo samples drawn from the posterior probability distribution (red dots), also for each simulation, as functions of the $SNR$ for the case of perfect wavelength calibration (left) and with simulated wavelength error (right). In the case of perfect wavelength calibration, only at low $\SNR$, in case of simulated wavelength error, both at low and high $\SNR$, obvious deviations of the empirical error from the asymptotic error are visible.

At low signal-to-noise, the empirical error diverges sharply from the asymptotic error, which could indicate either that the optimization algorithm did not find the real maximum of posterior, or that the logarithm of the posterior is no longer approximated well by a quadratic function around its maximum. However, the error estimates determined from the covariance matrix of the Monte Carlo samples are consistent with the empirical error at low $\SNR$ which indicates that the maximum finder worked as expected but instead, the posterior becomes non-Gaussian around the mode. To investigate the Gaussianity of the posterior, we plot the skewness and kurtosis of the Monte Carlo samples drawn from the Bayesian posterior in panels (c)-(f) of Figure~\ref{fig:rv_err_methods}. These indicate that the posterior -  or more precisely its marginalization over all parameters except $\vlos$ -  is indeed Gaussian at higher $\SNR$, but can become non-Gaussian at low signal-to-noise values.


At high signal-to-noise, when perfect wavelength calibration is assumed, the empirical error and the asymptotic error remain consistent and follow the same $(\SNR)^{-1}$ scaling as at intermediate signal-to-noise. This result is unrealistic as the practical error floor of fiber-fed spectrographs is thought to be around $1/100$ of a pixel which is equivalent to $0.6$~km~s$^{-1}$ for the MR arm. On the other hand, the result shows, that despite the pixelization of the spectrum, Doppler shift measurements could theoretically be better than the $1/100$ pixel limit if there was a way to more precisely calibrate the instruments. When we include the effect of incorrect wavelength calibration, the empirical error calculated from multiple simulations shows the expected error floor, as visible in the top right panel of Figure~\ref{fig:rv_err_methods}. Neither the asymptotic error, nor the error from Monte Carlo sampling, is capable of accounting for the error floor, however. This is the result of the simplification of the likelihood function which only accounts for the uncertainty of the observed flux but has no additional error terms to account for the imprecise wavelengths. We will address this issue in an upcoming paper.

\subsection{Effect of mismatched template parameters}
\label{sec:template_mismatch}

So far we have seen how the uncertainty of line-of-sight velocity measurement depends on the signal-to-noise ratio of the observed spectrum when fitting templates with atmospheric parameters that match the input parameters of the simulated spectra. Let us now study the case when the template parameters are far off the real values. This is a realistic scenario because synthetic spectra are pre-computed on grids of the parameters and the atmospheric parameters of observed stars seldom coincide with the grid points. On the other hand, we do not take into account the imperfections of the models because we generated the simulated observations from the same synthetic spectrum grid from which we take our templates.

\begin{figure}
    \begin{center}
        \includegraphics{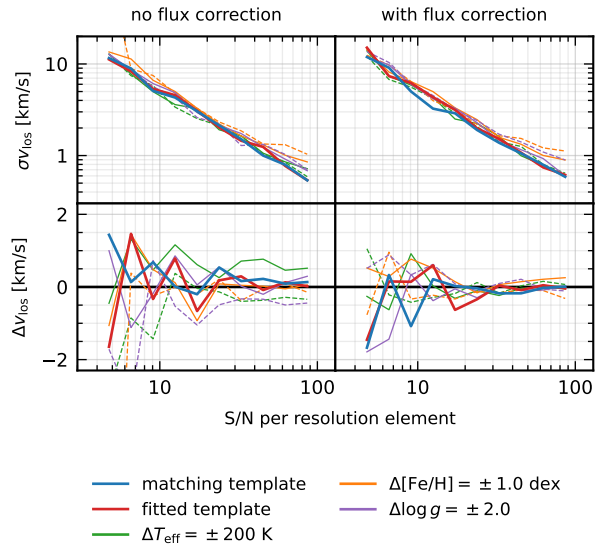}
        \caption{The variance and bias of line-of-sight velocity fits to simulations of a star with atmospheric parameters $\MH = -1.5$, $\Teff = 4750$~K and $\logg = 2.0$, with and without flux correction. The statistics are calculated from an ensemble of simulations in logarithmic bins of $\SNR$. The blue curves indicate the results from fitting a matching template with the same parameters as the input model, whereas the red curves show the result from fits where we optimized for the atmospheric parameters as well. The rest of the curves, as indicated by the color given in the legend, show the results when we used templates that are one or two grid points away from the input model, on the synthetic stellar spectrum grid. Solid (dashed) lines indicate a deviation in the positive (negative) direction from the input parameters for $\MH$, $\Teff$ and $\logg$. See text for detailed discussion.}
        \label{fig:rv_bias}
    \end{center}
\end{figure}

In Figure~\ref{fig:rv_bias}, we plot the uncertainty and bias of $\vlos$ as a function of $\SNR$ determined using template spectra with different atmospheric parameters. The curves show the standard deviation (top panels) and the mean (bottom panels) of $\Delta \vlos$ for many repeated simulations. The observations were simulated with artificially introduced fluxing errors, as described in Section~\ref{sec:simulation_details} and we repeated fitting $\vlos$ with and without correcting for fluxing errors. The tests were performed on simulated observations of a model with atmospheric parameters $\MH = -1.5$, $\Teff = 4750$~K and $\logg = 2.0$, whereas the templates were chosen to be one or two grid points away in the direction of each parameter (but only one parameter at a time). Step sizes of $\Delta \Teff = \pm 200$~K, $\Delta \MH = \pm 1.0$~dex and $\Delta \logg = \pm 2.0$ were used in effective temperature, metallicity and surface gravity, respectively.

The two top panels of Figure~\ref{fig:rv_bias} show $\sigma{\vlos}$ from fitting the different templates with and without the polynomial flux correction. The curves overlap remarkably well which indicates that using mismatched templates have very little effect on the uncertainty of $\vlos$. In general, fitting the simulated spectra with a template that matches the input atmospheric parameters, or letting the parameters vary freely yield the best results in terms of bias and error variance while non-matching templates tend to cause a bias of $1$-$2$~km~s$^{-1}$, even at higher $\SNR$, which is consistent with earlier results \citep{Walker2015MNRAS.448.2717W}. Templates with mismatched metallicity have the largest detrimental effect on the precision of $\vlos$ measurements.

Slight deviations from the exactly matching template (blue curves) are visible only at $\SNR > 50$ where using certain templates resulted in an overestimate of the uncertainty. However, as we showed in Section~\ref{sec:error_estimators} and illustrated in Figure~\ref{fig:rv_err_methods}, at such high signal-to-noise, errors of the wavelength calibration already start to dominate the uncertainty which we did not take into account in the simulations.

A stronger effect on the bias of $\vlos$ is visible in the two bottom panels of Figure~\ref{fig:rv_err_methods} when using mismatched templates to fit the line-of-sight velocity. Using a polynomial for flux correction tends to eliminate the bias, as it can be seen in the bottom right panel of Figure~\ref{fig:rv_err_methods}. It is important to point out, however, that even the bias from fitting with a perfectly matching template (blue curves) shows visible fluctuation at lower $\SNR$ that we attribute to the small number of simulations.

\subsection{Correlations of the template parameter \\ errors with $\sigma(\vlos)$}

\begin{figure}
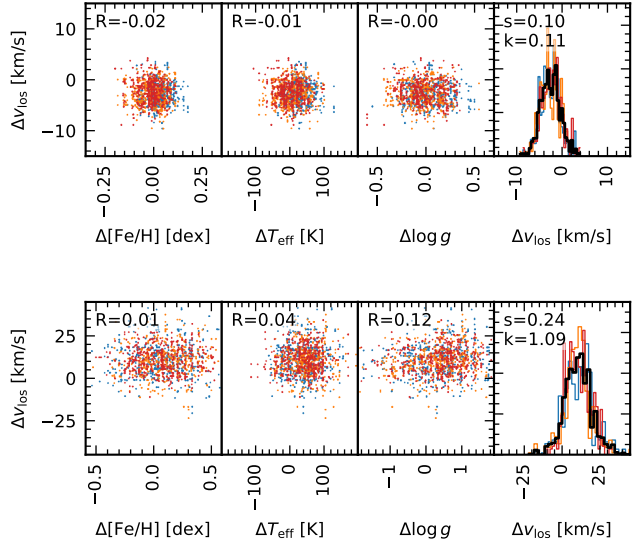

    \begin{center}
        \includegraphics{fig\_mcmc.pdf} \\
        \includegraphics{fig\_mcmc\_lowsnr.pdf}
        \caption{Results from Monte Carlo sampling of the posterior for two different realizations of the model with atmospheric parameters $\MH = -1.5$, $\Teff = 4750$~K and $\logg = 2.0$, and signal-to-noise (measured per resolution element in the MR arm) of $\SNR = 11.76$ (top) and $\SNR = 2.33$ (bottom).  The first three panels of each row show scatter plots of the deviations of the atmospheric parameters from the model input versus $\Delta \vlos$. The colors indicate  samples from three different Monte Carlo chains. The rightmost panel of each row shows the histograms of $\Delta \vlos$ for each Monte Carlo chain in color, whereas the black histogram is the combined histogram of all samples. In the top left of each scatter plot, the correlation coefficient of the two variables is printed. The small values of Pearson correlation indicate that $\vlos$ does not correlate with the atmospheric parameters. In the corner of the histograms, the values of the skewness $s$ and the kurtosis $k$ (zero for a Gaussian distribution) of the Monte Carlo samples are indicated. Note the different scales on the axes.}
        \label{fig:mcmc}
    \end{center}
\end{figure}

In order to assess the correlations between the measurement error of $\vlos$ and the variance of atmospheric parameters of the best matching template spectrum, we fitted the simulated spectra with full Monte Carlo sampling of the posterior probability. To illustrate the results of a Monte Carlo run, we used scatter plots of $\Delta \vlos$ as a function of the difference of atmospheric parameters from their input values in Figure~\ref{fig:mcmc} for a single simulated observation of the model with atmospheric parameters $\MH = -1.5$, $\Teff = 4750$~K and $\logg = 2.0$. The various colors show three different Monte Carlo chains, with 1000 samples each, that are plotted without any thinning. The good mixing of the Monte Carlo chains is apparent, as well as that practically no correlation between the atmospheric parameters and $\vlos$ exists. For reference, the Pearson correlation coefficients are printed in the top left corner of each panel.

\begin{figure}
    \begin{center}
        \includegraphics{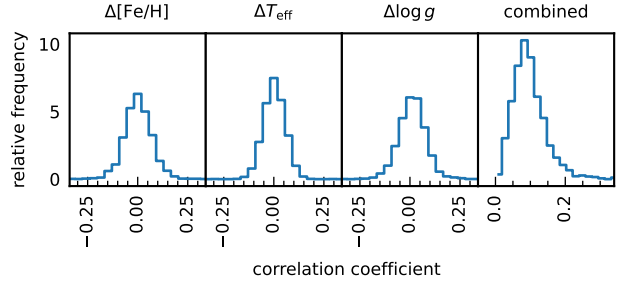}
        \caption{The distribution of the correlation coefficient of $\vlos$ and the atmospheric parameters (left three panels) and the coefficient of multiple correlation (rightmost panel), calculated from the Monte Carlo sampling of all realizations of all spectral types. According to these, the covariance of $\vlos$ with the template parameters is minimal on average.}
        \label{fig:mcmc_rv_corr}
    \end{center}
\end{figure}

In Figure~\ref{fig:mcmc_rv_corr}, we plot the histogram of correlation coefficients between $\vlos$ and the error of the template parameters for all realizations of all models in Tab.~\ref{tab:model_params} to demonstrate that the correlations are indeed very small, independently of stellar type and $\SNR$. We also plot the distribution of the coefficient of multiple correlation, which expresses how well $\vlos$ can be expressed as a linear combination of the atmospheric parameters within a Monte Carlo sample drawn from the posterior. Similar to the bivariate correlation, a value much smaller than $1$ means linear independence. The bivariate correlations are normally distributed around $0$ with a standard deviation of about $0.1$ which indicates the lack of any significant correlations. This reinforces the results presented in Section~\ref{sec:template_mismatch}, where we observed that fitting the simulated spectra with templates that did not precisely matched the input parameters did not result in increased uncertainties of the line-of-sight velocity. This is true in a broad range of signal-to-noise except for high $\SNR$ where the effect of using a mismatched template is more clearly detectable. On the other hand, at high $\SNR$, our simulations are inaccurate because we do not take any kind of wavelength calibration error or systematics into account.

\section{Discussion}
\label{sec:discussion}

\subsection{Scaling relations of $\sigma(\vlos)$ and $\SNR$}

To estimate the uncertainty of line-of-sight velocity measurements from the signal-to-noise ratio of observations and the primary parameters of the spectrograph arms, \cite{HC1992ESOC...40..275H} introduced the formula
\begin{equation}
    \sigma^\mathrm{HC}(\vlos) = 1.45 \times 10^6 \cdot (S/N)^{-1} \cdot R^{-1} \cdot B^{-1/2},
    \label{eq:hatzes_cochran}
\end{equation}
where velocity is measured in km~s$^{-1}$, $S/N$ is the signal-to-noise ratio of the flux measurement per pixel\footnote{The exact definition of $S/N$ is unspecified by \cite{HC1992ESOC...40..275H}. Hence, we assume it was set to the same in every pixel of their simulations.}, $R$ is the dimensionless spectral resolution and $B$ is the total wavelength coverage in $\Angstrom$. While the formula is certainly an approximation -- since it does not take important effects into account, such as the varying signal-to-noise ratio over the wavelength coverage, nor the contrast, width and density of the absorption lines -- it captures the primary relation between the uncertainty of line-of-sight velocity measurements and the square root of $S/N$.

\cite{Bouchy2001A&A...374..733B} pointed out that the information a spectral pixel carries about Doppler shift depends on the slope of the original spectrum (after instrumental broadening) in the pixel. They provide the formula
\begin{equation}
    \sigma^\mathrm{BPQ}(\vlos) = c \cdot \frac{1}{\sqrt{\sum W_p}}, 
    \quad
    \mathrm{where}
    \quad
    W_p = \frac{\lambda_p^2 \left[ \partial f_p / \partial \lambda \right]^2}{\sigma^2(f_p)}
    \label{eq:bouchy}
\end{equation}
to predict the measurement error of $\vlos$. On the other hand, when the $\left[ \partial f_p / \partial \lambda \right]^2$ slope is to be calculated numerically at pixel boundaries from the neighboring spectral pixels, one has to take the covariances into account, so the symbol $\sum W_p$ becomes
\begin{equation}
    \sum W_p = \mathrm{Tr}\,\, \mathbf{w}^\intercal \mathbf{C}^{-1} \mathbf{w},
\end{equation}
where the $\mathbf{w}$ vector is one element shorter than $f_p$ and takes the form of
\begin{equation}
    w_i = \frac{1}{2} (\lambda_i + \lambda_{i + 1}) 
          \left[ \frac{f_{i + 1} - f_i}{\lambda_{i + 1} - \lambda_i} \right].
\end{equation}
The covariance matrix is tridiagonal and can be written as
\begin{equation}
    C_{ij} =
    \begin{cases}
        \sigma^2(f_i) + \sigma^2(f_{i + 1}) & \mathrm{if} \,\, i = j \\
        \sigma^2(f_i)                       & \mathrm{if} \,\, j = i + 1 \,\, \mathrm{or} \,\, i = j + 1.
    \end{cases}
\end{equation}

Here we introduce a new quantity, the \textit{effective line number}, which can be calculated from synthetic spectra without identifying the individual lines. We are going to use the effective line number to derive a correction to Equation~\ref{eq:hatzes_cochran} of \cite{HC1992ESOC...40..275H}. Let us define the effective line number as
\begin{equation}
    \Neff = \sum \left| \frac{d f_i}{d \ln ( \lambda_i / 1~\Angstrom ) } \right| \bigg/
            \sum f_i
    \label{eq:n_eff}
\end{equation}
where $f_i$ is the flux in the $i^\mathrm{th}$ pixel and $\lambda_i$ is the wavelength measured at the pixel center. We use the derivative by $\ln \lambda$ since the broadening of spectral lines scales with the wavelength. This definition is additive in the sense that the effective number of lines scales with the wavelength coverage and the effective number of lines is always the sum of parts of the spectrum. We also introduce the quantity $\neff$, which is the effective number of lines \textit{per resolution element}.

\begin{figure}
    \begin{center}
        \includegraphics[]{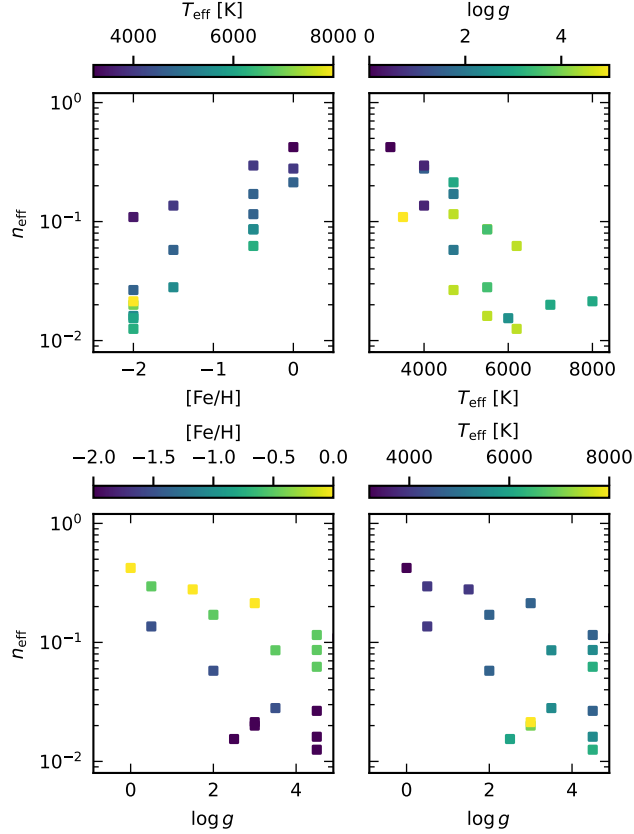}
        \caption{Dependence of the effective line number $\Neff$ on the atmospheric parameters in four different projections, calculated for the medium resolution arm MR. In this wavelength range, the effective number of lines depends primarily on metallicity and temperature and only secondarily on surface gravity, as it can be seen in the top two panels and and bottom left one, respectively. While the depth of hydrogen lines grows with temperature in the range we tested, the roughly inverse relation between $\Teff$ and $\Neff$ suggests that the calcium and iron lines play a more significant role.}
        \label{fig:n_eff_params}
    \end{center}
\end{figure}

Figure~\ref{fig:n_eff_params} illustrates the behavior of $\Neff$ as a function of the fundamental atmospheric parameters for the models in Tab.~\ref{tab:model_params}, where the effective line number is determined for the MR arm configuration. The strong dependence of the effective line number on metallicity is obvious, as well as that $\Neff$ decreases sharply with increasing temperature, at least in the parameter range of our models. This suggests that $\Neff$ is indeed sensitive to metal lines instead of the hydrogen line series. $\Neff$ also depends strongly on surface gravity, having a much larger value for giant stars than dwarf stars of the same metallicity. This is in accordance with our intuition that the line-of-sight velocity of giant stars with narrow lines can be measured with lower uncertainty at the same signal-to-noise ratio of the flux than dwarf stars with pressure-broadened lines.

To adopt Equation~\ref{eq:hatzes_cochran} to different stellar types, we calculated the ratio of the empirical $\sigma(\vlos)$, as measured from our simulations, and $\sigma_\textrm{HC}(\vlos)$, as calculated from Equation~\ref{eq:hatzes_cochran}, as a function of $\SNR$ for the arm configurations B and MR. The $95$~percentile of the per pixel signal-to-noise ratio turned out to be a much better characterization of noise in the continuum pixels than the median of the per pixel $\SNR$ when the overall $\SNR$ of spectra from different spectral types as compared. In Figure~\ref{fig:n_eff_Q}, we plot the ratio $\sigma(\vlos) / \sigma_\textrm{HC}(\vlos)$ as a function of the effective line count per resolution element. We find it remarkable how the points distribute along a single line in the log-log plot even though we did the analysis on fluxed spectra covering a broad range of atmospheric parameters.

\begin{figure}
    \begin{center}
        \includegraphics[]{fig\_rv_err_bouchy_hc.pdf}
        \caption{Comparison of $\sigma(\vlos)$ predictors of \protect\cite{HC1992ESOC...40..275H} (dashed black line, see Equation~\ref{eq:hatzes_cochran}) and \protect\cite{Bouchy2001A&A...374..733B} (solid black line, see Equation~\ref{eq:bouchy}) compared to our results from repeated simulations (blue lines) for a low metallicity K~giant with $\Teff = 4000$~K. The multiple blue lines indicate different logarithmic binning of $\Delta\vlos$ to calculate $\sigma(\vlos)$. The red line is a linear fit to the logarithm of the data with a slope of $-0.95$, very close to the theoretical value of $-1$.}
        \label{fig:rv_err_bouchy_hc}
    \end{center}
\end{figure}

\begin{figure}
    \begin{center}
        \includegraphics[]{fig\_n_eff_Q.pdf}
        \caption{The dependence of the ratio of Equation~\ref{eq:hatzes_cochran} of \cite{HC1992ESOC...40..275H} to empirical measurements of $\sigma(\vlos)$ based on our simulations, plotted as a function of the effective line number per resolution element as defined in Equation~\ref{eq:n_eff} for the arm configurations B (blue dots), R (red dots) and MR (black dots). Correcting the formula in Equation~\ref{eq:hatzes_cochran} using $\neff$ can account for a factor of 10 change in the uncertainty of line-of-sight velocity as a function of spectral type. The black line is not a fit but rather the function $0.6 \cdot \neff^{-3/4}$. The outlier point is an M giant, see Section~\ref{sec:M_giant} for discussion.}
        \label{fig:n_eff_Q}
    \end{center}
\end{figure}

The black line in Figure~\ref{fig:n_eff_Q} is not a fit to the data but indicates the function $0.6 \cdot \neff^{-3/4}$ that we can use to adapt Equation~\ref{eq:hatzes_cochran} to different spectral types in the form
\begin{equation}
    \sigma(\vlos) = 0.87 \times 10^6 \cdot \neff^{-3/4} \cdot (S/N)^{-1} \cdot R^{-1} \cdot B^{-1/2},
    \label{eq:hatzes_cochran_Neff}
\end{equation}
where the unit of velocity is km~s$^{-1}$.

\subsection{Different scaling of $\sigma(\vlos)$ for M giants}
\label{sec:M_giant}

In the bottom right panel of Figure~\ref{fig:rv_error}, we plotted the uncertainty of $\vlos$ as a function of the 95 percentile $\SNR$ per resolution element for various combinations of the spectrograph arms. The scaling of $\sigma(\vlos)$ is significantly different for the medium resolution red arms as the other two and also differs from the scaling of $\sigma(\vlos)$ of any other models. The flattening of the black curve is most likely due to the adopted definition of an overall $\SNR$, which characterizes the rest of the studied stellar types well.

In Figure~\ref{fig:snr_methods}, we plotted the per pixel $\SNR$ for three different stellar types in the MR arm, as a function of wavelength, where the top panel is for an M~giant, compared to the normalized $\SNR$ of a metal poor M dwarf and a metal poor K giant. The horizontal red line indicates the 95~percentile of $\SNR$ which we used on the x-axis of Figure~\ref{fig:rv_error}. The 95~percentile was chosen to define a measure of noise which is mostly sensitive to the continuum pixels where the $\SNR$ is the best -- in case of absorption spectra. Figure~\ref{fig:snr_methods} clearly shows that the mean and median $\SNR$ cannot capture the continuum pixels while the 95~percentile works for most stellar types except when the absorption features are so strong that we see no continuum pixels anymore. As a consequence, the M giant is an outlier on the rest of the plots as well.

\subsection{Uncertainties at low $\SNR$ from MCMC}

We have seen in Section~\ref{sec:error_estimators} that the uncertainty estimate of $\vlos$ calculated from the Fisher matrix becomes unreliable at low $\SNR$ and significantly underestimates the empirical error determined from repeated simulations. On the other hand, Monte Carlo sampling of the same objective function yields uncertainties consistent with the empirical errors. Even when flux correction coefficients are set aside, a full Monte Carlo sampling of the posterior of $\vlos$ and the template parameters has a high computational cost compared to evaluating the Fisher matrix at the best fit parameters.

Nevertheless, some applications might benefit from measuring $\vlos$ from a large number of stellar spectra observed at very low $\SNR$. For example, the surface density of relatively bright stars of faint satellite galaxies of the Milky Way might be too low to fill the fibers of an instrument similar to the Subaru PFS. However, measuring the line-of-sight velocity of a large number of fainter stars -- even if only with high uncertainties -- might carry useful information when the uncertainties are estimated correctly and the data is treated in a statistically robust way.

\section{Conclusions}
\label{sec:conclusions}

We have investigated the dependence of the uncertainty of line-of-sight velocity measurements on the signal-to-noise ratio of the flux in the case of a wide selection of stellar types that are relevant to Galactic Archaeological observations. We found that at intermediate $\SNR$, the uncertainty of $\vlos$ scales with the square root of $\SNR$ as expected from theory, but significant deviations arise at low $\SNR$. We have shown that asymptotic estimators of the uncertainty, such as the Cram{\' e}r--Rao lower bound, calculated from the Fisher matrix, is unable to correctly characterize $\sigma(\vlos)$ when measured from noisy spectra, but a full Monte Carlo sampling of the posterior probability can account for the empirical variance that we see in repeated simulations of the template fitting.

We have calculated the elements of the Fisher matrix for the parameters of the best fitting template spectrum and also given analytic results for the uncertainty of the line-of-sight velocity. We have shown how the full covariance matrix of the flux correction function coefficients and the template parameters can be calculated efficiently when the systematics of flux calibration of the observation or the differences between the template spectrum and the observation require correcting the stellar continuum with some wavelength dependent correction function in the form of a linear combination of base functions, such as polynomials.

We have defined a new quantity, the effective absorption line density, which can be used to estimate the expected uncertainty of line-of-sight velocity measurements from a spectrum with a certain stellar type, as a function of continuum signal-to-noise. The method works as well as the uncertainties calculated by \cite{Bouchy2001A&A...374..733B} and gives more realistic results than the formula of \cite{HC1992ESOC...40..275H}.

Our work informs best practices for estimating uncertainties in $\vlos$.  Our conclusions are relevant to the new era of massively multiplexed spectroscopy, featuring spectrographs such as Subaru PFS, DESI, WEAVE, 4MOST, and VLT MOONS\@. These spectrographs will inevitably observe many -- perhaps a majority -- stars at low $\SNR$ where we have found that simple estimators of uncertainty are inaccurate. We advocate for the computation of full posterior $\vlos$ distributions for science cases where accurate estimates of uncertainty are important, such as the computation of dark matter density profiles.

\begin{acknowledgments}
This work is supported by the generosity of Eric and Wendy Schmidt, by recommendation of the Schmidt Futures program.
E.N.K.\ acknowledges support from NSF CAREER grant AST-2233781.
\end{acknowledgments}

\appendix

\section{Properties of the Fisher matrix}
\label{ax:fisher}

\subsection{Numerical Evaluation Scheme for the Fisher matrix}
\label{ax:fisher_numerical}

While Equation~\ref{eq:fisher_simple} is an intuitive analytic result, this is not the most convenient way to compute the Fisher matrix numerically. In case of high $\SNR$, values of the significance function $\nu$ become large, and in the second derivative, one needs to take small differences of large numbers, leading to fairly large numerical errors. We can minimize the errors by calculating the relevant second derivatives as pixel-wise sums over the numerical derivatives of the template spectrum. In order to facilitate this, we will introduce the following set of new functions containing the derivatives of the models.
\begin{eqnarray}
    \psi_{00} = \sum_p \dfrac{m_p(z_0) m_p(z_0)}{\sigma_p^2}\\
    \psi_{01} = \sum_p \dfrac{m_p(z_0) m_p'(z_0)}{\sigma_p^2}\\
    \psi_{11} = \sum_p \dfrac{m_p'(z_0) m_p'(z_0)}{\sigma_p^2}\\
    \psi_{02} = \sum_p \dfrac{m_p(z_0) m_p''(z_0)}{\sigma_p^2}\\
    \phi_{00} = \sum_p \dfrac{f_p m_p(z_0)}{\sigma_p^2}\\
    \phi_{01} = \sum_p \dfrac{f_p m_p'(z_0)}{\sigma_p^2}\\
    \phi_{02} = \sum_p \dfrac{f_p m_p''(z_0)}{\sigma_p^2}
\end{eqnarray}
It turns out that the function $\psi_{02}$ will never appear in the Fisher matrix. We can also define the matrix $\Psi$ as
\begin{equation}
    \Psi = \begin{pmatrix}
        \psi_{00}& \psi_{01}\\
        \psi_{01} & \psi_{11}
    \end{pmatrix} 
\end{equation}
Also note, that $\psi_{00} = \chi(z_0)=\chi_0$. The ensemble averages of the different terms taken at the maximum of the significance function are the following.
\begin{equation}
    \begin{aligned}
        \langle\varphi\rangle_0 &=A_0 \phi_{00}, &
        \langle\varphi'\rangle_0 &=A_0 \phi_{01}, &
        \langle\varphi''\rangle_0 &=A_0 \phi_{02}, \\
        \langle\chi\rangle_0 &=\psi_{00}, &
        \langle\chi'\rangle_0 &=2\,\psi_{01}, &    
        \langle\chi''\rangle_0 &=2\,(\psi_{11}+\psi_{02}).
    \end{aligned}
\end{equation}
At the maximum of the significance function $\phi_{00}=A_0 \psi_{00}$ and $2 \phi_{01}/\phi_{00} = \psi_{01}/\psi_{00}$ holds. We can use this and substitute the ensemble averages of the derivatives into the curvature matrix at the true position of the maximum, giving us the Fisher matrix in the form of
\begin{equation}
    \bF = 
    \begin{pmatrix}
        \psi_{00}  &  A_0\psi_{01} \\[12pt]
        A_0\psi_{01} &  
        A_0^2(\psi_{11}+\psi_{02})-A_0\phi_{02}
    \end{pmatrix}
\end{equation}
We can write the determinant of $F$ as
\begin{equation}
    \mdet{\bF} = A_0^2 \mdet{\Psi} + A_0 \psi_{00} (A_0 \psi_{02}-\varphi_{02}).
\end{equation}
The last term in the determinant should go to zero as we approach the optimal template. Now we can compute the $\sigma_z^2$ element, which gives the Cram\'er-Rao lower bound on the error of line-of-sight velocity measurement as
\begin{equation}
    \sigma_z^2 = \dfrac{\psi_{00}}
                       { \left| \bF \right| }
\end{equation}
We can use $A_0 = \varphi_0/\chi_0$ to reduce this to quantities that are already calculated during the likelihood evaluation. In any case, $A_0$ is dimensionless, $\chi_0$ scales with the inverse variance, $\Psi$ scales as the square of the inverse variance. Overall, $\sigma_z^2$ should approximately scale with the variance of the noise, and its units should be km~s$^{-1}$ from $|\Psi|$.

\subsection{Including the template parameters}
\label{ax:fisher_model_parameters}

Including the template parameters into the optimization -- as opposed to only optimize for the line-of-sight velocity with a fixed template -- involves taking the partial derivatives of the significance function with respect to the atmospheric parameters $\theta_\alpha$. The curvature matrix of the likelihood surface will be larger, augmented by a row and column for each parameter, but the expressions for the matrix elements will be very similar. Here we will use the subscript $z$ to denote differentiation with respect to $z$, and $\alpha$ and $\beta$ to denote differentiation with respect to the template parameters. Omitting the zero subscript for the maximum, the augmented Fisher matrix can be written as
\begin{equation}
    \mdet{\bF} = 
    \begin{pmatrix}
        \chi &
            \varphi_1 &
                \varphi_\beta\\[7pt] \\
        \varphi_1 &
            -\nu\nu_{11} +\dfrac{{\varphi_1}^2}{\chi} &
                    -\nu\nu_{1\beta}+\dfrac{\varphi_1\varphi_\beta}{\chi}\\[14pt]
        \varphi_\alpha &
            -\nu\nu_{\alpha 1} +\dfrac{{\varphi_\alpha\varphi_1}}{\chi} &
                -\nu\nu_{\alpha\beta} +\dfrac{\varphi_\alpha\varphi_\beta}{\chi}
    \end{pmatrix}.
\end{equation}
In order to calculate the Cram\'er--Rao bound for the Doppler shift, we need to calculate the $zz$ matrix element of the inverse matrix. This requires calculating the determinant of the matrix 
\begin{equation}
    \bS = 
    \begin{pmatrix}
        \chi &
            \varphi_\beta \\[14pt]
        \varphi_\alpha &
            -\nu\nu_{\alpha\beta} +\dfrac{\varphi_\alpha\varphi_\beta}{\chi}
    \end{pmatrix},
\end{equation}
where we eliminated the column and row of $\bF$ corresponding to the $zz$ element. The velocity error bound $\sigma_z^2$ is given by
\begin{equation}
    \sigma_z^2 = \dfrac{\mdet{\bS}}{\mdet{\bF}}.
\end{equation}
We can get a relatively simple expression if we re-write these matrices in a block form as
\begin{equation}
    \bF = 
    \begin{pmatrix}
        \bA   &  \bB\\[7pt]
        \bB^\T & \bD
    \end{pmatrix}  
    \quad \mathrm{and} \quad
    \bS = 
    \begin{pmatrix}
        \ba   &  \bb\\[7pt]
        \bb^\T & \bd
    \end{pmatrix},
\end{equation}
where $\bB$ and $\bb$ are row vectors. The determinants of $\bF$ and $\bS$ can be expressed as
\begin{align}
    \mdet{\bF} & = \mdet{\bA} \mdet{\bD - \bB^\T \bA^{-1} \bB} =
        \mdet{\bA} \mdet{\bG} \\
    \mdet{\bS} & = \mdet{\ba} \mdet{\bd - \bb^\T \ba^{-1} \bb} =
        \mdet{\ba}\mdet{\bg},
\end{align}
where we introduced $\bG = \bD - \bB^\T \bA^{-1} \bB$ and $\bg = \bd - \bb^\T \ba^{-1} \bb$.

Let us now consider our actual problem where, in the optimum of the likelihood function, the matrix blocks will be $\bA=\chi$, $\bB = (\varphi_z, \varphi_\alpha)$, and $\bA^{-1} = 1/\chi$, thus
\begin{equation}
    G_{ij} = -\nu\nu_{ij},
\end{equation}
where $i,j=(z,\alpha...)$ go over $z$ and the atmospheric parameters of the template. Similarly,
\begin{equation}
    g_{ij} = -\nu\nu_{\alpha\beta},
\end{equation}
but the indices only run over the templates parameters $\alpha$. Here $g$ is the Hessian of the log-likelihood at the maximum point, with respect to the template parameters, and its determinant is the Gaussian curvature. With these transformations, the variance of the Doppler shift becomes much simpler. With $\mdet{\bA} = \mdet{\ba} = \chi$, it can be written as
\begin{equation}
    \sigma_z^2 = \dfrac{\mdet{\bg}}{\mdet{\bG}}.
\end{equation}
Let us separate off the first row and column of $\bG$ and write it as
\begin{equation}
    \bG = 
    \begin{pmatrix}
        -\nu\nu_{zz} & -\nu\nu_{z\beta }\\[7pt]
        -\nu\nu_{\alpha z} & g_{\alpha\beta}
    \end{pmatrix}.
\end{equation}
Knowing that $G_{11} = -\nu \nu_{zz}$, the determinant of $\bG$ can be expressed as
\begin{equation}
    \mdet{\bG} = 
        -\nu \nu_{zz}
            \bigmdet{\bg + \dfrac{\nu\nu_\alpha\nu_\beta}{\nu_{zz}}} =
        -\nu \nu_{zz} |\,\hat \bg\,|,
\end{equation}
where
\begin{equation}
    \hat g_{\alpha\beta} = g_{\alpha\beta} + \left(\dfrac{\nu}{\nu_{zz}}\right)\nu_\alpha\nu_\beta
\end{equation}
with which the variance becomes
\begin{equation}
    \sigma_z^2 = \left( \dfrac{1}{-\nu\nu_{11}} \right)
                 \dfrac{\mdet{\bg}}{\mdet{\hat\bg}}.
\end{equation}
One can see how this this is different from Equation~\ref{eq:z_error_from_nu}, where the template parameters were not considered. We will now show that $\mdet{\bg} \,/\, \mdet{\hat\bg} \leq 1$ which indicates that the error of the line-of-sight velocity measurement decreases when the template parameters are optimized as well.

Let us factorize the symmetric, positive definite matrix $\bg$ as $\bg = \bU^\T \bSigma \bU$, where $\bSigma$ is diagonal and rotate the $\nu_\gamma$ partial derivatives of the significance with $\bU$ as
\begin{equation}
    s_\beta = \nu_\gamma U_{\gamma\beta}, \qquad\qquad  s^\T_\alpha =
        U_{\alpha\gamma}^\T \nu_\gamma .
\end{equation}
Since the determinant is invariant under the rotation, we can write $\mdet{\hat \bg}$ as
\begin{equation}
    \mdet{\hat \bg} =
    \bigmdet{
        \bU^\T \bSigma\, \bU +
            \left( \dfrac{\nu}{\nu_{zz}} \right)
            \bU^\T \bs^\T \bs\, \bU
    } = 
    \bigmdet{
        \bSigma + \left(\dfrac{\nu}{\nu_{zz}}\right) \bs^\T \bs
    }
\end{equation}
According to the determinant lemma, see Appendix~\ref{sec:det_lemma}, the determinant of a matrix of this special form (diagonal plus an outer product of a vector with itself) can be calculated analytically as
\begin{equation}
    \mdet{\hat\bg} = 
        \mdet{\bg} \left( 1 + \sum_\alpha \dfrac{s_\alpha^2}{\lambda_\alpha} \right),
\end{equation}
where $\lambda_\alpha$ are the eigenvalues (diagonal elements) of $\bSigma$.
This yields the very intuitive result of
\begin{equation}
    \sigma_z^2  =    \left(\dfrac{1}{-\nu\nu_{zz}}\right) \dfrac{1}{1 + \sum_\alpha \dfrac{s_\alpha^2}{\lambda_\alpha}},
\end{equation}
that relates to the simple case where we only maximized the likelihood to get the amplitude and the line-of-sight velocity, but not the atmospheric parameters. Furthermore, it shows that the more parameters we optimize for, the lower the uncertainty of the inferred line-of-sight velocity will be.

\subsection{Linear Model for Envelope Correction}
\label{ax:fluxcorr}

We made the assumptions in Sections~\ref{sec:significance_model_params}~and~\ref{ax:fisher_model_parameters} that the observed spectrum is noisy and the template differs from the observation only by a linear amplitude which assumption is obviously too simple. It is reasonable to expect that the models need to be corrected with a smooth multiplicative function of wavelength to account for small errors in spectrophotometric calibration, or slight differences in the continuum of the template and the real spectrum. As introduced in Section~\ref{sec:flux_correction}, we model the flux correction with a function that is a linear combination of functions $q_n(\lambda)$, such as Chebyshev or Legendre polynomials, that only depend on the wavelength:
\begin{equation}
    A(\lambda) = \sum_{n=0}^N A_n q_n(\lambda).
\end{equation}
The orthogonality of the functions is not necessary but the $q_n(\lambda_p)$ vectors must be linearly independent to ensure that we can always solve for the best fit $A_n$ coefficients when all template parameters, including $z$ are given.

We can now rewrite the log-likelihood using a vector notation, where the quantities $\bA$, $\bphi$ and $\bchi$ are vectors and matrices of dimension of $N + 1$, as
\begin{equation}
        \mathcal{L}(\bA,z) = \bA^\T \! \bphi -\frac{1}{2} \bA^\T \! \bchi \bA,
\end{equation}
where $\bphi(z)$ is a vector-valued, and $\bchi(z)$ is matrix valued function of the Doppler shift, defined as
\begin{align}
    \varphi_k(z) & = \sum_p q_k(\lambda_p) \frac{f_p m_p(z)} {\sigma_p^2}, \\
    \chi_{kn}(z) & = \sum_p q_k(\lambda_p) \, q_n(\lambda_p) \, \frac{m_p^2(z)}{\sigma_p^2},
\end{align}
These are the natural extensions of Equation~\ref{eq:phi_chi} from Section~\ref{sec:significance}, with the addition of the array of multiplicative functions $q_k$. Evaluating the partial derivatives of $\mathcal{L}$ with respect to $A_k$ and $z$, an equating them to zero, we get
\begin{align}
    \frac{\partial \mathcal{L}}{\partial \bA} & = \bphi(z) - \bchi(z) \bA = 0 
    \label{eq:likelihood_diff_fluxcorr}\\
    \frac{\partial \mathcal{L}}{\partial z} & =  \bA^\T \! \bphi'(z) 
        -\frac{1}{2} \bA^\T \! \bchi'(z) \bA = 0,
    \label{eq:likelihood_diff_z}
\end{align}
where we denoted the differentiation with respect to $z$ by the prime and continued to use vector notation. Since $\bchi$ is a symmetric, positive definite matrix, it has an inverse, and from the first equation we can solve for $\bA$ at the maximum point of $\mathcal{L}$, denoted by the zero subscript. The solution of Equation~\ref{eq:likelihood_diff_fluxcorr} is simply
\begin{equation}\label{eq:A0}
    \bA_0 = \bchi_0^{-1} \bphi_0, \qquad\qquad \bA_0^\T =\bphi_0^\T \bchi_0^{-1}.
\end{equation}
The solution of Equation~\ref{eq:likelihood_diff_z} gives us the line-of-sight velocity. At the maximum point with respect to the $A_n$ coefficients, it becomes
\begin{equation}
    \bA_0^\T \bphi_0' - \frac{1}{2} \bA_0^\T \bchi_0' \bA_0 = 
    \bA_0^\T \Big[\bphi_0' - \frac{1}{2}\bchi_0' \bA_0\Big] = 0,
\end{equation}
which means, that the quantity in the square brackets, composed of the first derivatives of $\bphi$ and $\bchi$, is orthogonal to $\bA_0$ at the maximum. Substituting the values of $\bA_0$ from Equation~\ref{eq:A0}, we can rewrite this in the form of
\begin{equation}
     \frac{\partial \mathcal{L}}{\partial z}\Bigg|_0 =
     \bphi_0^\T \bchi_0^{-1} \bphi_0' - \frac{1}{2} \bphi_0^\T \bchi_0^{-1} \bchi_0' \bchi_0^{-1} \bphi_0
     \label{eq:likelihood_diff_z_}
\end{equation}
Here and later we will be using the identity
\begin{equation}
    \bK' = -\bK \bchi_0' \bK, \qquad {\rm and}\qquad  \bK'' = -\bK \bchi_0''\bK + 2 \bK \bchi_0' \bK \bchi_0' \bK
\end{equation} 
for the derivative of the inverse of a symmetric invertible matrix, where $K=\bchiI$. With this, we can rewrite Equation~\ref{eq:likelihood_diff_z_} expression in the more suggestive form of
\begin{equation}
     \frac{\partial \mathcal{L}}{\partial z}\Bigg|_0 =
     \bphi_0^\T \bchi_0^{-1} \bphi_0' + \frac{1}{2} \bphi_0^\T (\bchi_0^{-1})' \bphi_0 = 
     \frac{\partial}{\partial z} \left(\frac{1}{2}  \bphi^\T \bchi{-1} \bphi \right)\Bigg|_0
     \end{equation}
The expression in the parentheses is identical in function to the significance function $\nu$, defined in Equation~\ref{eq:significance} of Section~\ref{sec:significance} as $\nu^2/2 = \varphi^2/\chi$. The significance function in vector form can be written as 
\begin{equation}
    \frac{\nu^2}{2} = \bphi^\T \bchi^{-1} \bphi,
\end{equation}
which can be evaluated at any Doppler shift $z$, and whose maximum coincides with the maximum of $\mathcal{L}$.

\subsection{The Fisher Matrix in Vector Form}

Similarly to Section~\ref{sec:significance}, we can calculate the elements of the Fisher information matrix. First, let us write down the second derivatives of the log likelihood, which are
\begin{align}
    \frac{\partial^2 \mathcal{L}}{\partial \bA^2}          & = -\bchi \\
    \frac{\partial^2 \mathcal{L}}{\partial \bA \partial z} & = \bphi' -\bchi' \bA \\
    \frac{\partial^2 \mathcal{L}}{\partial z^2}            & = \bA^\T \bphi''-\frac{1}{2}\bA^\T \bchi'' \bA.
\end{align}
The Fisher matrix is the negative of this, evaluated at the maximum point:
\begin{equation}
    F  = 
        \begin{pmatrix}
            \bchi_0& -\bphi_0' + \bchi_0' \bA_0\\[7pt]
            -\bphi_0'^\T+\bA_0^\T\bchi_0' & -\bA_0^\T \bphi_0''+\frac{1}{2}\bA_0^\T \bchi_0'' \bA_0
        \end{pmatrix}.
\end{equation}
This is a symmetric block matrix of size $N+2$, composed of an $N+1$ matrix $\bchi_0$ augmented by a single row and column that corresponds to the Doppler shift. We can use the so called \textit{bordering formula}, see Equation~\ref{eq:bordering}, to evaluate the elements of the inverse of the Fisher matrix $S=F^{-1}$, block by block, starting with $S_{22}$ as
\begin{align}
    \frac{1}{S_{22}} =
    & -\bA_0^\T \bphi_0''+\frac{1}{2}\bA_0^\T \bchi_0'' \bA_0
        - (\bphi_0'^\T-\bA_0^\T\bchi_0')\bchiI(\bphi_0'-\bchi_0' \bA_0) \\
    = 
    & -\bA_0^\T \bphi_0''+\frac{1}{2}\bA_0^\T \bchi_0'' \bA_0
        - \bphi_0'^\T \bchiI \bphi_0' -\bA_0^\T\bchi_0'\bchiI\bchi_0' \bA_0
        + \bA_0^\T\bchi_0'\bchiI\bphi_0',
\end{align}
where we already aggregated the symmetric terms. Let us now show, that this term is exactly $-\nu_0 \nu_0''$ by using the expressions from 
Equations~\ref{eq:K1},~\ref{eq:K2}~and~\ref{eq:A0}.
\begin{align}
    \left. \frac{\partial^2}{\partial z^2}\left(\frac{1}{2}\nu^2\right) \right|_0
        & = \left. \frac{1}{2} \left( \bphi^\T \bchi^{-1} \bphi\right)'' \right|_0 \\
        & = 
            \bphi_0^\T \bK_0 \bphi_0'' +\frac{1}{2}\bphi_0^\T\bK_0''\bphi_0 +\bphi_0'^\T\bK_0\bphi_0'
            + 2\bphi_0^\T\bK_0'\bphi_0' \\
        & =
            \bA_0^\T\bchi_0''-\frac{1}{2}\bA_0^\T\bchi_0''\bA_0+\bphi_0'^\T \bchiI \bphi_0'
            + \bA_0^\T\bchi_0'\bchiI\bchi_0'\bA_0-2 \bA_0^\T\bchi_0'\bchiI\bphi_0'
\end{align}
We can thus express $S_{22}$ through $\nu_0\nu_0''$ in a much simpler form as
\begin{equation}
    \mu = \frac{1}{S_{22}} = -\nu_0\nu_0'',
\end{equation}
and the rest of the matrix elements are
\begin{align}
    S_{11} = & \bchiI + \frac{1}{\mu}\bchiI 
        \left( -\bphi_0' + \bchi_0' \bA_0 \right)
        \left(-\bphi_0'^\T+\bA_0^\T\bchi_0' \right) \bchiI \\
    S_{12} = & -\bchiI \left( -\bphi_0' + \bchi_0' \bA_0 \right) \Big/ \mu \\
    S_{21} = & -\left(-\bphi_0'^\T+\bA_0^\T\bchi_0' \right) \bchiI \Big/ \mu
\end{align}
Next, we define the vector $\bB$ for the first derivatives as
\begin{equation}
    \bB_0 = \bchiI (\bphi_0'-\bchi_0' \bA_0).
\end{equation}
With this, the expression of inverse of the Fisher matrix simplifies considerably to
\begin{equation}
    \bf{S} = \begin{pmatrix}
        \bchiI-\dfrac{\bB_0 \bB_0^\T}{\nu_0\nu_0''} & \dfrac{\bB_0}{\nu_0\nu_0''} \\[14pt]
        \dfrac{\bB_0^\T}{\nu_0\nu_0''} & -\dfrac{1}{\nu_0\nu_0''}
    \end{pmatrix}.
\end{equation}
To us the most important is the $S_{22}$ element. This is in exact agreement with the previous, scalar result, except that $\nu^2$ is calculated as a quadratic form of $\bphi$ and $\bchi^{-1}$. The rest of the inverse matrix can be calculated numerically using the formulae provided above.

\section{Matrix identities}

In the paper, we use several identities to aid in the inversion of matrices and element-wise derivatives of matrices. Here we summarize these identities along with some explanation.

\subsection{The Bordering Formula}

To analytically calculate the inverse of the Fisher matrix, we used a matrix identity called the \textit{bordering formula}. Here $\mathbf{F}$ is a symmetric matrix of size $N+1$, built by padding an existing invertible square matrix $\bM$ of size $N$ with an extra row and column in 
a symmetric fashion, using the vector $\bd$ and the scalar $c$ as
\begin{equation}
    \bf{F}^{-1} = \begin{pmatrix}
        \bM & \bd \\[10pt]
        \bd^T & c
    \end{pmatrix}^{-1} =
    \begin{pmatrix}
        \bMi + \bMi\bd\bd^T\bMi/\mu & -\bMi \bd/\mu \\[10pt]
        -\bd^T \bMi /\mu & 1/\mu
    \end{pmatrix},
    \label{eq:bordering}
\end{equation}
where 
\begin{equation}
    \mu = c - \bd^T \bMi \bd.
\end{equation}

\subsection{Derivative of an inverse matrix}

This formula deals with computing the element-wise derivative of a matrix inverse. Let us take a symmetric invertible matrix $\bchi$, and denote its inverse as $\bK=\bchi^{-1}$. We can show that the derivative of $\bK$ can be calculated as $\bK' = -\bK \bchi' \bK$. Let us start from the equality
\begin{equation}
    \mathbf{1} = \bchi \bchi^{-1} = \bchi \bK.
\end{equation}
By differentiating the above equation we arrive at
\begin{equation}
    \mathbf{0} = \bchi' \bK + \bchi \bK'.
\end{equation}
Rearranging this gives
\begin{equation}\label{eq:K1}
     \bK' = -\bK \bchi' \bK,
\end{equation}
from which we can calculate the second derivative as
\begin{equation}\label{eq:K2}
     \bK'' = -2\bK'\bchi'\bK -\bK\bchi''\bK = -\bK\bchi''\bK +2\bK\bchi'\bK\bchi'\bK.
\end{equation}

\subsection{The determinant lemma}
\label{sec:det_lemma}

The determinant lemma states that the determinant of a matrix in a special form (diagonal plus an outer product of a vector with itself) can be calculated as
\begin{equation}
    \mdet{\bA \mathbf{u}\mathbf{v}^\T} = \mdet{\bA} \left( 1 + \mathbf{v}^\T \bA^{-1} \mathbf{u} \right),
\end{equation}
where $\mathbf{v}$ and $\mathbf{u}$ are arbitrary column vectors.

\bibliography{main}{}
\bibliographystyle{aasjournal}

\end{document}